\documentclass[11pt,a4paper]{article}
\usepackage{jheppub}
\usepackage{dsfont}
\usepackage[vcentermath]{youngtab}

\usepackage{amsmath, amssymb}								
\usepackage{amsmath, amssymb,wasysym}
\usepackage{mathrsfs}

\usepackage{datetime}

\newcommand{\be}{ \begin{equation}}
\newcommand{\ee}{\end{equation}}
\newcommand{\bea}[1]{\begin{eqnarray}\label{#1} }
\newcommand{\eea}{\end{eqnarray}}

\newcommand{\ch}{{\rm ch}}

\DeclareMathOperator{\su}{\mathfrak{su}}
\DeclareMathOperator{\tr}{tr}

\newcommand{\mc}{\mathcal }
\newcommand{\mk}{\mathfrak}
\newcommand{\mb}{\mathbb }
\newcommand{\ba}{\begin{eqnarray}}
\newcommand{\ea}{\end{eqnarray}}

\title{The large $\mathcal{N}=4$ superconformal $\mathcal{W}_\infty$ algebra}

\author[a]{Matteo Beccaria,}
\author[b]{Constantin Candu,}
\author[b]{and Matthias R.~Gaberdiel}

\affiliation[a]{Dipartimento di Matematica e Fisica ìEnnio De Giorgiî,\\
Universitaí del Salento \& INFN, Via Arnesano, 73100 Lecce, 
Italy}
\affiliation[b]{Institut f\"ur Theoretische Physik,\\ 
ETH Zurich, CH-8093 Z\"urich, Switzerland}

\emailAdd{matteo.beccaria@le.infn.it}
\emailAdd{canduc@itp.phys.ethz.ch}
\emailAdd{gaberdiel@itp.phys.ethz.ch}

\abstract{
The most general large ${\cal N}=4$ superconformal ${\cal W}_{\infty}$ algebra, containing in addition to the 
superconformal algebra one supermultiplet for each integer spin, is analysed in detail. It is found that the
${\cal W}_{\infty}$ algebra is uniquely determined by the levels of the two $\su(2)$ algebras, a
conclusion that holds both for the linear and the non-linear case. We also perform various cross-checks 
of our analysis, and exhibit two different types of truncations in some detail.}

\allowdisplaybreaks
\begin{document}

\maketitle
\flushbottom

\section{Introduction}

The duality between higher spin theories on AdS$_3$ \cite{Vasiliev:1995dn,Vasiliev:1999ba}
and  large $N$ limits of $2$d CFTs, see
\cite{Gaberdiel:2012uj} for a review, can be understood and tested in quite some detail. This
applies, in particular, to the bosonic example of \cite{Gaberdiel:2010pz}, thus suggesting that 
supersymmetry is not a crucial ingredient for 
these types of dualities.  On the other hand, it is believed that the
vector-like higher spin/CFT dualities arise from a full
stringy AdS/CFT correspondence upon taking the tensionless limit and concentrating on the states
belonging to the leading Regge trajectory 
\cite{Sundborg:2000wp,Witten,Mikhailov:2002bp}. In this context the supersymmetric versions
of the dualities  naturally arise, and thus the supersymmetric examples deserve 
special attention.
There have been some attempts to understand in detail
the way in which the higher spin/CFT dualities fit into string theory, 
see e.g.~\cite{Sagnotti:2011qp} for a review as well as the proposal in \cite{Chang:2012kt}; however, 
it is fair to say that there are still many open questions. The $3$d/$2$d case seems to be a very promising
arena to explore these issues in more detail since both sides of the duality are under very good quantitive control. 

With this vision in mind, the analysis of the ${\cal N}=4$ supersymmetric  version of the higher spin/CFT duality 
was initiated in \cite{Gaberdiel:2013vva}. It relates the higher spin theory based on the Lie algebra
$\mathfrak{shs}_2[\lambda]$ to the Wolf space cosets 
\begin{equation}
\frac{\mathfrak{su}^{(1)}(N+2)_{k+N+2}}{\mathfrak{su}(N)^{(1)}_{N+k+2}\oplus\mathfrak{u}(1)} 
\qquad \hbox{with} \qquad \lambda = \frac{N}{N+k+2} \ .  \label{eq:cosets}
\end{equation}
These theories have `large' ${\cal N}=4$ 
superconformal symmetry, which is the expected superconformal symmetry of the 
dual to string theory on ${\rm AdS}_3\times {\rm S}^3 \times {\rm S}^3 \times {\rm S}^1$. In a sense
this case is more restrictive than the better explored ${\rm AdS}_{3} \times {\rm S}^3 \times M_4$ case
with $M_4=\mathbb{T}^4$ or $M_4={\rm K3}$, in which case only the small ${\cal N}=4$ superconformal
algebra is expected to appear. In particular, the large ${\cal N}=4$ superconformal algebra contains
two affine $\su(2)$ algebras, and the small ${\cal N}=4$ superconformal algebra can be obtained 
as a contraction in the limit in which one of the levels is sent to infinity. The other reason for 
studying the case with large ${\cal N}=4$ superconformal symmetry is that the dual CFT of string
theory is unknown \cite{Gukov:2004ym} (see however \cite{Tong:2014yna} for a recent proposal), and 
one may hope that the novel 
higher spin perspective may also suggest new avenues for overcoming this impasse. Finally, it would
be very interesting to make contact with the approach based on the integrable spin chain
viewpoint of \cite{OhlssonSax:2011ms,Borsato:2012ss}. 

The proposal of  \cite{Gaberdiel:2013vva} was subsequently explored further. In particular, the spectrum of the 
two descriptions was matched in \cite{Candu:2013fta}, see also \cite{Creutzig:2013tja} for an earlier analysis,
and the asymptotic symmetry algebra of the higher spin theory was shown to agree with the 't~Hooft limit of the 
Wolf space coset ${\cal W}$ algebras \cite{Gaberdiel:2014yla}. While many of the features of this duality
mirror precisely what happens for the original bosonic proposal \cite{Gaberdiel:2010pz} and its ${\cal N}=2$
supersymmetric generalisation \cite{Creutzig:2011fe,Candu:2012jq}, there is one intriguing difference
that was already noticed in \cite{Gaberdiel:2013vva}: while the quantum ${\cal W}_\infty$ algebras
underlying the bosonic and the ${\cal N}=2$ version exhibit a triality or quadrality relation 
\cite{triality,Candu:2012tr}, respectively, that explains the identification of the quantisation 
of the asymptotic symmetry algebra with the dual coset algebra even at finite $N$, a similar
relation does not seem to exist in the large ${\cal N}=4$ case. It is therefore interesting to 
understand the structure of the  large ${\cal N}=4$  quantum ${\cal W}_{\infty}$ algebra in detail. 

This is what will be done in this paper. As we shall see,  the relevant quantum ${\cal W}_{\infty}$ algebra is 
uniquely determined in terms of the levels of the two affine $\mathfrak{su}(2)$ algebras.
As a consequence, the quantisation of the asymptotic symmetry algebra of the higher spin theory 
must coincide with the coset algebra provided that the levels of the two $\mathfrak{su}(2)$ algebras agree,
thus explaining the agreement of the symmetries without a triality-like relation. The absence of 
such a relation only implies that the 
quantisation of the asymptotic symmetry algebra of the higher spin theory based 
on the finite dimensional higher spin algebra $\mathfrak{shs}_2[\lambda]$ with $\lambda=M$ integer is {\em not} isomorphic 
to the Wolf space coset (\ref{eq:cosets}) with  $N=M$. In fact, as we shall also explain in detail, 
while both algebras truncate to some finitely 
generated quantum algebras at integer $M$, the precise structure of the truncation is rather different in the two 
cases.
\medskip

The paper is organised as follows. In Section~2 and 3 we study the structure of the non-linear
large ${\cal N}=4$ ${\cal W}_\infty$ algebra. In particular, we explain our conventions for 
the supermultiplets in Section~2, and  make the most general ansatz for the various OPEs in Section~3.2. 
We then study the constraints that follow from imposing the associativity of the OPEs, and describe
our results in Section~3.3 (as well as Appendix~B). In Section~3.4 we analyse the different truncation
patterns of this ${\cal W}_\infty$ algebra, and explain how the finitely generated symmetry algebras
associated to $\mathfrak{shs}_2[M]$ and the coset algebra at finite $N$, respectively, fit into this picture. In 
Section~4 we repeat the analysis for the case of the linear ${\cal N}=4$ ${\cal W}_{\infty}$ algebra,
and find essentially the same structure. As a non-trivial consistency check of our analysis we explain
in detail in Section~4.3 and 4.4 how the two sets of results are related to one another upon going from the 
linear to the non-linear description. Section~5 contains our conclusions, and some of the more technical
material has been relegated to three appendices.


\section{The Non-linear Large $\mathcal{N}=4$ Superconformal Algebra}\label{sec:nl4}


In this section we explain our conventions for the description of the large ${\cal N}=4$ superconformal
algebra, its superprimaries and  their descendants.

\subsection{The OPEs of the Superconformal Algebra}

The non-linear large $\mathcal{N}=4$ superconformal algebra is generated by the stress energy tensor
\begin{equation}
T(z)T(w)\sim \frac{c}{2(z-w)^4}+\frac{2T(w)}{(z-w)^2}+\frac{\partial T(w)}{z-w}\ ,
\label{eq:TTans}
\end{equation}
six spin $1$ currents $A^{\pm i}$, $i=1,2,3$, which are primary with respect to $T$ and generate an 
$\su(2)_{k_+}\oplus \su(2)_{k_-}$ subalgebra 
\begin{equation}
A^{\pm i}(z)A^{\pm j}(w)\sim \frac{k_\pm \eta^{ij}}{(z-w)^2}+\frac{{f^{ij}}_l A^{\pm l}(w)}{z-w} \ ,
\label{eq:su2}
\end{equation}
as well as four spin $\frac{3}{2}$ supercharges $G^{\alpha\beta}$ which are primary with respect to both $T$ and 
the currents $A^{\pm i}$
\begin{equation}
A^{+ i}(z)G^{\alpha\beta}(w)\sim \frac{\rho^i_{\gamma\alpha}G^{\gamma\beta}}{z-w}\ ,\qquad  \ A^{- i}(z)G^{\alpha\beta}(w)\sim \frac{\rho^i_{\gamma\beta} G^{\alpha\gamma}}{z-w}\ .
\end{equation}
Here $\rho^i$ denotes the spin $j=\frac{1}{2}$ representation of $\su(2)$,
and the $\su(2)$ invariant bilinear form $\eta$ in eq.~\eqref{eq:su2} is defined by 
$\eta^{ij}= \tr \rho^i \rho^j$.
Global $\su(2)\oplus\su(2)$ symmetry constrains the  OPEs of the supercharges to take the following most general quadratic form
\begin{multline}
G^{\alpha\beta}(z)G^{\gamma\delta}(w)\sim \frac{b \epsilon^{\alpha\gamma}\epsilon^{\beta\delta}}{(z-w)^3}
+\left[\frac{1}{(z-w)^2}+\frac{\partial}{2(z-w)}\right]\left(s^+ \epsilon_{\beta\delta}\ell_{i,\alpha\gamma} A^{+i}
+s^- \epsilon_{\alpha\gamma}\ell_{i,\beta\delta} A^{-i}\right)(w) \\ 
\quad + \frac{1}{z-w} \left[\epsilon_{\alpha\gamma}\epsilon_{\beta\delta}(-4T + s^{++} \eta_{ij}A^{+i }A^{+j}
+s^{--}\eta_{ij}A^{-i }A^{-j} )+ s^{+-}\ell_{i,\alpha\gamma}\ell_{j,\beta\delta} A^{+i}A^{-j}\right](w)\ ,
\label{eq:GGans}
\end{multline}
where $\epsilon_{\alpha\beta}$ is the antisymmetric matrix with $\epsilon_{12}=1$, and the matrices $\ell_i$ are defined by
\begin{equation}
\ell_{i,\alpha\beta} = \epsilon_{\alpha\gamma}\,  \rho^j_{\gamma\beta}\,  \eta_{ji}\ .
\end{equation}
Here  $\eta_{ij}$ is the inverse of $\eta^{ij}$, and in the following we shall routinely use these two matrices to raise 
and lower the indices in the adjoint representation. 
The Jacobi identities fix the structure constants  in eqs.~(\ref{eq:TTans}) and (\ref{eq:GGans})
to \cite{goddard} (see also 
\cite{Fradkin:1992km})
\begin{align}
c&=\frac{3(k_++k_-+2k_+k_-)}{k_++k_-+2}=\frac{6(k_++1)(k_-+1)}{k_++k_-+2}-3 \ ,
& b &= -\frac{8k_+k_-}{2+k_+ + k_-}\ ,\\
 s^\pm &= \frac{8 k_\mp}{2+k_++k_-}\ ,\quad \qquad s^{\pm\pm}= \frac{2}{2+k_++k_-}\ ,& s^{+-}&= -\frac{8}{2+k_++k_-}\ .
\end{align}
In the limit $k_\pm \to \infty$ with the ratio
\begin{equation}
\alpha = \frac{k_-}{k_+} \qquad \hbox{kept fixed,}
\end{equation}
the wedge modes of the the non-linear large $\mathcal{N}=4$ superconformal algebra generate
 the exceptional Lie superalgebra  $D(2,1;\alpha)$. Conversely, the non-linear large $\mathcal{N}=4$
 superconformal algebra can be constructed as the Drinfel'd-Sokolov reduction of $D(2,1;\alpha)$ \cite{henneaux}.


\subsection{Superprimaries and their Descendants}\label{sec:sprim}


We call a field ${\cal N}=4$ superprimary provided that it is primary with respect to the 
stress-energy tensor $T$, as well as the currents $A^{\pm i}$. In addition, we require that the 
OPEs with the supercharges $G^{\alpha\beta}$ only have first order poles; in terms of the corresponding
state these
conditions are equivalent to requiring that it is annihilated by the positive modes
of the stress-energy tensor, the currents and the supercharges, respectively.

In general, an ${\cal N}=4$ superprimary then transforms in an (irreducible) representation of 
the zero modes $A^{\pm i}_0$  of $\su(2)\oplus \su(2)$; in the following we shall consider the 
case where this representation is the singlet representation.
We then denote the superconformal descendants of the superconformal primary $V^{(s)}$ by
%
\begin{equation}
\begin{array}{c|c|c|c|c|c}
\text{component } & V^{(s)}_0 & V^{(s)\alpha\beta}_{1/2} & V^{(s)\pm i}_{1} & V^{(s)\alpha\beta}_{3/2} & V^{(s)}_2\\ \hline
\text{conformal spin } & s & s+1/2 & s+1 & s+3/2 & s+2\\ 
\text{$\mathfrak{su}(2)\oplus\mathfrak{su}(2)$  spin }& (0,0) & (1/2,1/2) & (1,0) \oplus (0,1) & (1/2,1/2) & (0,0)\\
\end{array}\ .
\label{eq:nonlinear-components}
\end{equation}
Here $s$ is the conformal dimension of the superprimary field $V^{(s)}_0$, and the structure of the
multiplet is as described in \cite{Gaberdiel:2013vva}, see also \cite{jeugt}.

The precise form of the OPEs of these component fields with the fields of the 
large $\mathcal{N}=4$ superconformal algebra depend, to a certain extent, on our
conventions.\footnote{This is to be contrasted with the case of the linear superconformal algebras where
requiring that the defining OPEs are linear usually leads to a unique choice. In the present case, a linear basis does
not exist, and we need to fix this ambiguity differently.} We have chosen to work with a quasiprimary
basis, and the guiding principle for our conventions has been to minimise the number of non-linear terms.
For example, for the OPEs of the component fields with the stress-energy tensor  we make the ansatz
\begin{align}
T(z)V^{(s)}_0(w)&\sim \frac{sV^{(s)}_0(w)}{(z-w)^2}+\frac{\partial V^{(s)}_0(w)}{z-w}\ , \notag\\
T(z)V^{(s)\alpha\beta}_{1/2}(w)&\sim \frac{(s+\frac{1}{2})V^{(s)\alpha\beta}_{1/2}(w)}{(z-w)^2}
+\frac{\partial V^{(s)\alpha\beta}_{1/2}(w)}{z-w}\ , \notag  \\
T(z)V^{(s)\pm i}_{1}(w)&\sim \frac{(s+1)V^{(s)\pm i}_{1}(w)}{(z-w)^2}+\frac{\partial V^{(s)\pm i}_{1}(w)}{z-w}\ ,\label{Tans}\\
T(z)V^{(s)}_{3/2}(w)&\sim \frac{(s+\frac{3}{2})V^{(s)\alpha\beta}_{3/2}(w)}{(z-w)^2}+\frac{\partial V^{(s)\alpha\beta}_{3/2}(w)}{z-w}\ ,
\notag \\
T(z)V^{(s)}_2(w)&\sim \frac{t V^{(s)}_0(w)}{(z-w)^4} + \frac{sV^{(s)}_2(w)}{(z-w)^2}+\frac{\partial V^{(s)}_2(w)}{z-w}\ .\notag
\end{align}
Note that these fields are Virasoro primary, except for $V^{(s)}_2$, which is only quasi-primary if $t\neq 0$ (as 
will be generically the case, see below). Similarly, as regards their behaviour under the current algebra, we postulate
\begin{align}
A^{\pm i}(z) V^{(s)}_0(w)&\sim 0\ , \notag\\
A^{+ i}(z) V^{(s)\alpha\beta}_{1/2}(w)&\sim \frac{\rho^i_{\gamma\alpha}V^{(s)\gamma\beta}_{1/2}(w)}{z-w}\ ,
\quad A^{- i}(z) V^{(s)\alpha\beta}_{1/2}(w)\sim \frac{\rho^i_{\gamma\beta}V^{(s)\alpha\gamma}_{1/2}(w)}{z-w}\ , \notag\\
A^{\pm i}(z) V^{(s)\pm j}_{1}(w)&\sim \frac{a^\pm _1\eta^{ij}V^{(s)}_0(w)}{(z-w)^2}+\frac{{f^{ij}}_l V^{(s)\pm l}_1(w)}{z-w}\ ,
\quad A^{\pm i }(z) V^{(s)\mp j}_{1}(w) \sim 0\ , \label{Aans}\\
A^{+ i}(z) V^{(s)\alpha\beta}_{3/2}(w)&\sim \frac{a^+_{3/2} \rho^i_{\gamma\alpha}V^{(s)\gamma\beta}_{1/2}(w)}{(z-w)^2}+ \frac{ \rho^i_{\gamma\alpha}V^{(s)\gamma\beta}_{3/2}(w)}{z-w}\ , \notag \\
A^{- i}(z) V^{(s)\alpha\beta}_{3/2}(w)&\sim \frac{a^-_{3/2}\rho^i_{\gamma\beta}V^{(s)\alpha\gamma}_{1/2}(w)}{(z-w)^2}+\frac{\rho^i_{\gamma\beta} V^{(s)\alpha\gamma}_{3/2}(w)}{z-w}\ , \notag \\
A^{\pm i}(z)V^{(s)}_2(w)&\sim \frac{a_{2}^\pm V^{(s)\pm i}_1(w)}{(z-w)^2}\ . \notag
\end{align}
Thus $V^{(s)}_0$ and $V^{(s)}_{1/2}$ are affine-primary, but the higher component fields are not (since there are
double poles in the OPEs with the currents). Our conventions for the OPEs with the supercharges
are given in appendix~\ref{app:supmult}, and the associativity of this ansatz with the ${\cal N}=4$ fields
then implies that we have to choose 
\be\label{c1}
t =  -\frac{48 s (1 + s) (k_+ - k_-)}{(1 + 2 s) (2 + k_+ + k_-)}\ ,  
\ee
\be\label{c2}
a_1^\pm = 4s\ ,\quad a^\pm_{3/2} = \pm \frac{8 (1 + s) [1 + k_\pm +s (2 +  k_+ +  k_-)]}{(1 + 2 s) (2 + k_+ + k_-)}\ ,
\quad a^\pm_{2}= \pm 4(1+s)\ ,
\ee
as well as the values given in eq.~(\ref{c3}).


\section{Non-linear Large $\mathcal{N}=4$ $\mathcal{W}_\infty$ algebra}


With these preparations we are now ready to study the structure of the ${\cal W}$ algebra that contains
in addition to the non-linear large $\mathcal{N}=4$ superconformal algebra higher spin multiplets
 $V^{(s)}$ of spin $s=1,2,3,\dots$ --- one multiplet for every positive integer spin.

We shall use the same methods as in \cite{triality, evenhol, Candu:2013uya, Beccaria:2013wqa}: first we write down 
the most general ansatz for the OPEs between the higher spin currents that are allowed by the basic requirements
of conformal symmetry. Then we impose the Jacobi identities to solve for the structure constants in these OPEs. 
Our primary goal is to understand how many non-equivalent such $\mathcal{W}_\infty$ algebras exist, i.e., 
whether there are any further free parameters, in addition to $k_\pm$, that characterise these algebras.

In this section we shall write all OPEs in a quasiprimary basis. The OPEs then take the general form \cite{wrep}
\begin{equation}
\Phi^i(z)\, \Phi^j(w)= \sum_{k}  \frac{{C^{ij}}_k}{(z-w)^{h_i+h_j-h_k}}  \sum_{n=0}^{\infty}  
\frac{(h_i-h_j+h_k)_n}{n!(2h_k)_n}\, (z-w)^n \, \partial^n \Phi^k(w) \ ,
\label{eq:opefull}
\end{equation}
where $\Phi^i$, $\Phi^j$, $\Phi^k$ are quasi-primary operators of conformal dimension $h_i$, $h_j$ and $h_k$, 
respectively, ${C^{ij}}_k$ are the structure constants and  $(x)_n=\Gamma(x+n)/\Gamma(x)$ denotes the 
Pochhammer symbol.
In order to improve the readability of the following formulas, we shall always use the 
shorthand notation for the singular part of the OPEs of type~\eqref{eq:opefull}
\begin{equation}
\Phi^i \times \Phi^j \sim \sum_{k\, :\, h_k < h_i+h_j}  {C^{ij}}_k \Phi^k\ .
\label{eq:opeshort}
\end{equation}
It should be obvious how to recover the actual
singular part of the OPE~\eqref{eq:opefull}  from the shorthand expression~\eqref{eq:opeshort}.

\subsection{Composite Fields}

In order to be able to write down the most general ansatz for the OPEs of the higher spin fields
in a quasiprimary basis we first need to find all the quasiprimary operators at every spin.
A convenient (albeit somewhat formal) way of doing this is as follows. We introduce a 
`mark'  for every field of the algebra
\begin{equation}
\begin{array}{c||c|c|c||c|c|c|c|c}
\text{component} & A^{\pm i}& G^{\alpha\beta}& T& V^{(s)}_0 & V^{(s)\alpha\beta}_{1/2} & V^{(s)\pm i}_{1} & V^{(s)\alpha\beta}_{3/2} & V^{(s)}_2\\ \hline
\text{mark} & y^\pm_{0,1}& y_{0,3/2} & y_{0,2}& y_{s,0} & y_{s,1/2} & y^\pm_{s,1} & y_{s,3/2} & y_{s,2}\\ 
\end{array}\ .
\end{equation}
Then, the marked character of the full $\mathcal{W}_\infty$ algebra takes the form
\begin{equation}
\chi_{\infty}= \chi_0 \cdot \chi_{\text{hs}}\ ,
\label{eq:chw}
\end{equation}
where $\chi_0$ is the character of the large ${\cal N}=4$ superconformal algebra
\begin{equation}
\chi_0 = \prod_{n=1}^\infty \frac{\prod^{\frac{1}{2}}_{m,m'=-\frac{1}{2}}(1+y_{0,\frac{3}{2}}z_+^{2m}z_-^{2m'}q^{n+\frac{1}{2}})}
{(1-y_{0,2}q^{n+1})
\prod^1_{m=-1}(1-y^+_{0,1}z_+^{2m}q^n)(1-y^-_{0,1}z_-^{2m}q^n)} \ ,
\label{eq:chn4}
\end{equation}
$z_\pm$ are the chemical potentials for the two $\su(2)$ algebras, 
and $\chi_{\text{hs}}$ counts the states generated by the higher spin fields
\begin{multline}
\chi_{\text{hs}}=  \prod_{s=1}^\infty\prod_{n=s}^\infty \frac{\prod^{\frac{1}{2}}_{m,m'=-\frac{1}{2}}(1+y_{s,\frac{1}{2}}z^{2m}_+ z^{2m'}_-q^{n+\frac{1}{2}})
(1+y_{s,\frac{3}{2}}z^{2m}_+ z^{2m'}_-q^{n+\frac{3}{2}})}{(1-y_{s,0}q^n)(1-y_{s,2}q^{n+2})\prod^1_{m=-1}(1-y^+_{s,1}z_+^{2m}q^{n+1})(1-y^-_{s,1}z^{2m}_-q^{n+1})}\ .
\label{eq:chhs}
\end{multline}
The quasiprimary fields at spin $s$ are then counted by the `multiplicities' $d_s$, where
\begin{equation}
\chi_\infty = 1+ \sum_{s\in\mathbb{N}/2}\frac{d_s q^s}{1-q}\ .
\end{equation}
The first few $d_s$ are explicitly 
\begin{align*}
d_1 ={}& y_{1,0}+y^+_{0,1}\ch_1(z_+) + y^-_{0,1}\ch_1(z_-)\ ,\\
d_{\frac{3}{2}} ={}& \big(y_{0,\frac{3}{2}}+y_{1,\frac{1}{2}}\big)\ch_{\frac{1}{2}}(z_+)\ch_{\frac{1}{2}}(z_-)\ ,\\
d_2= {}&[y_{0,2} + y_{2,0}+ (y_{1,0})^2+(y^+_{0,1})^2+(y^-_{0,1})^2\big] + \big(y^+_{1,1}+y^+_{0,1}y_{1,0}\big)\ch_1(z_+)+{}\\
&{}+(y^-_{1,1}+y^-_{0,1}y_{1,0})\ch_1(z_-)+ y^+_{1,1}y^-_{1,1}\ch_1(z_+)\ch_1(z_-)\ ,\\
d_{\frac{5}{2}}={}& \big[y_{1,\frac{3}{2}}+y_{2,\frac{1}{2}}+(y_{1,0}+y^+_{0,1}+y^-_{0,1})(y_{0,\frac{3}{2}}+y_{1,\frac{1}{2}})\big]\ch_{\frac{1}{2}}(z_+)\ch_{\frac{1}{2}}(z_-)+{}\\
&{}+y^+_{0,1}(y_{0,\frac{3}{2}}+y_{1,\frac{1}{2}})\ch_{\frac{3}{2}}(z_+)ch_{\frac{1}{2}}(z_-) + y^-_{0,1}(y_{0,\frac{3}{2}}+y_{1,\frac{1}{2}})\ch_{\frac{1}{2}}(z_+)\ch_{\frac{3}{2}}(z_-)\ ,\\
d_3 = {}& \big\{
y_{3,0} + y_{1,2} + y_{1,0}\big [y_{0,2} + y_{2,0}+ (y_{1,0})^2+(y^+_{0,1})^2+(y^-_{0,1})^2
\big] + y^+_{0,1}y^+_{1,1}+y^-_{0,1}y^-_{1,1} +{}\\
&{}+ y_{0,\frac{3}{2}}y_{1,\frac{1}{2}} \big\} +
\big\{
y^+_{2,1} + y_{1,0}\big(y^+_{1,1}+y^+_{0,1}+y_{1,0} y^+_{0,1}\big) + 
y^+_{0,1}[y_{0,2}+y_{2,0}+ y^+_{0,1}+y^+_{1,1} +{}\\
&{}+ ( y^+_{0,1})^2  +( y^-_{0,1})^2] + y_{0,\frac{3}{2}}+y_{0,\frac{3}{2}}y_{1,\frac{1}{2}}+(y_{1,\frac{1}{2}})^2
\big\}\ch_1(z_+)+ \big\{
y^-_{2,1} + y_{1,0}\big(y^-_{1,1}+y^-_{0,1}+{}\\
&{}+y_{1,0} y^-_{0,1}\big) + 
y^-_{0,1}[y_{0,2}+y_{2,0}+ y^-_{0,1}+y^-_{1,1} + ( y^-_{0,1})^2  +( y^+_{0,1})^2] + y_{0,\frac{3}{2}}+y_{0,\frac{3}{2}}y_{1,\frac{1}{2}}+{}\\
&{}+(y_{1,\frac{1}{2}})^2
\big\}\ch_1(z_-)+ \big[ 
y_{0,\frac{3}{2}}y_{1,\frac{1}{2}} + y^+_{1,1}y^-_{0,1} + y^+_{0,1}y^-_{1,1} + y^+_{0,1}y^-_{0,1} y_{1,0}+ y^+_{0,1}y^-_{0,1}
\big] \times {}\\
&{}\times \ch_1(z_+)\ch_1(z_-) + \big[ y^+_{1,1}y^+_{0,1}+(y^+_{0,1})^2 y_{1,0}\big]\ch_2(z_+) + \big[ y^-_{1,1}y^-_{0,1}+(y^-_{0,1})^2 y_{1,0}\big]\times {}\\
&{}\times \ch_2(z_-) + (y^+_{0,1})^3  \ch_3(z_+)+(y^-_{0,1})^3  \ch_3(z_-)+
(y^+_{0,1})^2 y^-_{0,1}  \ch_2(z_+)\ch_1(z_-)+{}\\
&{}+(y^-_{0,1})^2 y^+_{0,1}  \ch_1(z_+)\ch_2(z_-)\ ,
\end{align*}
where $\ch_j(z)= \sum^j_{m=-j} z^{2m}$  is the  character of the $\mathfrak{su}(2)$ representation of spin $j$.
{}From the explicit expressions for $d_s$ we can verify that all quasiprimaries up to spin $3$ are given by
\begin{align*}
s = 1:&\quad
V^{(1)}_0 \ ,\ A^{\pm i}\ ,\\
s= 3/2:&\quad 
G^{\alpha\beta}\ ,\ V^{(1)\alpha\beta}_{1/2}\ ,\\
s= 2:&\quad 
T\ , V^{(2)}_0\ ,\ V^{(1)\pm i}_1\ , \ [V^{(1)}_0V^{(1)}_0]\ ,\ [A^{\pm i} V^{(1)}_0]\ ,\ [A^{\pm i }A^{\pm j }]\ ,\ [A^{+ i }A^{- j }] \ ,\\
s = 5/2:&\quad
V^{(1)\alpha\beta}_{3/2}\ ,\ V^{(2)\alpha\beta}_{1/2}\ ,\ [V^{(1)}_{0}V^{(1)\alpha\beta}_{1/2}]\ ,\ [V^{(1)}_{0} G^{\alpha\beta}]\ ,\ [A^{\pm i}V^{(1)\alpha\beta}_{1/2}]\ ,\ [A^{\pm i}G^{\alpha\beta}]\ ,\\
s = 3:&\quad
V^{(3)}_0 \ ,\ V^{(2)\pm i}_1 \ ,\ V^{(1)}_2 \ ,\ [V^{(1)}_0 V^{(2)}_0]\ ,\ [V^{(1)}_0 V^{(1)\pm i}_1]\ ,\ [V^{(1)}_0 [V^{(1)}_0 V^{(1)}_0]]\ ,\\
 &\quad [V^{(1)\alpha\beta}_{1/2}V^{(1)\gamma\delta}_{1/2}]\ ,\ [G^{\alpha\beta}V^{(1)\gamma\delta}_{1/2}]\ ,\ [G^{\alpha\beta}G^{\gamma\delta}]\ ,\ [A^{\pm i} V^{(2)}_0]\ ,\
 [A^{\pm i} V^{(1)\pm j}_1]\ ,\\
&\quad  [A^{\pm i} V^{(1)\mp j}_1]\ ,\ [A^{\pm i} V^{(1)}_0]_{-1}\ ,\ [A^{\pm i} [V^{(1)}_0V^{(1)}_0]]\ ,\ [A^{\pm i} [A^{\pm j} V^{(1)}_0]]\ ,\\
&\quad  [A^{+ i} [A^{- j} V^{(1)}_0]]\ ,\ [T V^{(1)}_0]\ ,\ [A^{\pm i}[A^{\pm j}A^{\pm l}]]\ ,\ [A^{\pm i}[A^{\pm j}A^{\mp l}]]\ ,\ [TA^{\pm i}]\ ,\\
&\quad  [A^{\pm i}A^{\pm j}]_{-1}\ ,\
[A^{+ i}A^{- j}]_{-1}\ .
\end{align*}
Here we have introduced a modified normal ordered product $[\Phi^i \Phi^j ]$, which is characterised
by the property that it defines a quasiprimary operator provided that $\Phi^i$ and $\Phi^j$ are quasiprimary.
More precisely, this modified normal ordered product differs from the standard normal ordered product $(\Phi^i \Phi^j )$ by the descendants of the quasiprimary operators appearing in the poles of the OPE~\eqref{eq:opefull}
\begin{equation}
(\Phi^i\Phi^j) = [\Phi^i\Phi^j]  + \sum_k \, {C^{ij}}_k \binom{2h_i-1}{h_i+h_j-h_k} \,
\frac{\Gamma(2h_k)}{\Gamma(h_i+h_j+h_k)}\, \partial^{h_i+h_j-h_k}\Phi^k\ .
\end{equation}
We have also introduced the following quasiprimary fields\footnote{These fields can be rewritten in terms of the normal ordered product $\mathscr{N}(\Phi^i,\partial^n \Phi^j)$ defined in  \cite{Blumenhagen:1990jv}.}
\begin{align*}
[A^{\pm i} V^{(1)}_0]_{-1} &= \frac{1}{2} (\partial A^{\pm i} V^{(1)}_0) - \frac{1}{2}( A^{\pm i} \partial V^{(1)}_0) \ ,\\
[A^{\pm i} A^{\pm j}]_{-1} &= \frac{1}{2} (\partial A^{\pm i} A^{\pm j}) - \frac{1}{2}( A^{\pm i} \partial A^{\pm j})  - \frac{1}{12} {f^{ij}}_l \partial ^2 A^{\pm l}\ ,\\
[A^{+ i} A^{- j}]_{-1} &=\frac{1}{2} (\partial A^{+ i} A^{- j}) - \frac{1}{2}( A^{+ i} \partial A^{- j})\ .
\end{align*}
\smallskip

We can also deduce from the marked character the number of (composite) $\mathcal{N}=4$ superprimary fields 
that transform in the singlet representation $(0;0)$ of $\su(2)\oplus \su(2)$ at spin $s$. To this end 
we expand the marked character with $y^\pm_{0,1}=y_{0,3/2}= y_{0,2}=1$ and 
$y_{s,0}=y_{s,1/2}=y^\pm_{s,1}=y_{s,3/2}=y_{s,2}=y_s$, in terms of characters of $\mathcal{N}=4$ superprimaries
%
%
\begin{equation}
\chi_\infty = \chi_0 + \sum_{s\in \mathbb{N}/2}  e_s q^s\times\prod_{n=1}^\infty
\frac{ \prod_{m,m'=-\frac{1}{2}}^{\frac{1}{2}} (1+z_+^{2m}z_-^{2m'}q^{n-\frac{1}{2}})}{(1-q^n)
\prod^1_{m=-1}(1-z_+^{2m}q^{n})(1-z^{2m}_-q^{n})}\ ,
\end{equation}
where $e_s$ is the `multiplicity' of the $\mathcal{N}=4$ superprimaries at spin $s$. We can further decompose 
$e_s$ into $\su(2)\oplus\su(2)$ characters to get the `multiplicity' of the superprimaries in a given representation
\begin{equation}
e_s = \sum_{l_+,l_-} e_{s}(l_+,l_-) \, \ch_{l_+}(z_+) \, \ch_{l_-}(z_-) \ .
\end{equation}
The first few values of $e_s(0,0)$ are then
\begin{align}\notag
e_1(0,0) &= y_1\ ,\\\notag
e_{2}(0,0) &= y_2 + y_1^2\ ,\\
e_{3}(0,0) &= y_3 + y_1^3+ y_1y_2 \ ,
\label{eq:n4comp}
\end{align}
and it is not hard to convince oneself that $e_s(0,0)=0$ for all half-integer values of $s$.
Thus, at spin $s=2$ there is a single composite superprimary of the form $[V^{(1)}_0V^{(1)}_0]+\cdots$, 
which can be used to redefine $V^{(2)}_0$, while at spin $s=3$ there are two composite superprimaries of the form
$[V^{(1)}_0[V^{(1)}_0V^{(1)}_0]]+\cdots$ and $[V^{(1)}_0V^{(2)}_0]+\cdots$, which can be used to redefine $V^{(3)}_0$.

\subsection{Ansatz for OPEs}\label{sec:ope}


With these preparations we can now make the most general ansatz for the OPEs between the various higher spin
fields (up to total spin $4$). Our ansatz will obviously need to respect the $\su(2)\oplus\su(2)$  symmetry (coming from the 
zero modes of the currents). At total spin $2$ and $\frac{5}{2}$, the most general ansatz is then
\begin{align}\label{eq:ope2}
V^{(1)}_0 \times V^{(1)}_0 \sim{}& n_1 I + 0\cdot V^{(1)}_0\ ,&
V^{(1)}_0 \times V^{(1)\alpha\beta}_{1/2} \sim{}&  w_1 G^{\alpha\beta}+ 0\cdot V^{(1)\alpha\beta}_{1/2}
\ .
\end{align}
Here the coefficient in front of $V^{(1)}_0$ vanishes because a single spin one current can only generate an abelian Kac-Moody algebra.
It is also clear that the coefficient in front of $V^{(1)\alpha\beta}_{1/2}$ must vanish because, by conformal
symmetry, the 3-point function 
\be
\langle V^{(1)}_0(z) \, V^{(1)\alpha\beta}_{1/2}(w)\,  V^{(1)\gamma\delta}_{1/2}(v)\rangle 
\ee
is symmetric under the exchange of $w$ and $v$, which however is incompatible with the fermionic
nature of these fields.
\smallskip

\noindent At total spin $3$ the most general ansatz for the OPEs is then
\begin{align}\notag
V^{(1)}_0 \times V^{(1)+ i}_{1} \sim{}&  w_2 A^{+ i} + w_3 [A^{+ i}V^{(1)}_0] + w_4 V^{(1)+ i}_1\ ,\\ \notag
V^{(1)}_0 \times V^{(1)- i}_{1} \sim{}&  w_5 A^{- i} + w_6 [A^{- i}V^{(1)}_0] + w_7 V^{(1)- i}_1\ , \\ \notag
V^{(1)\alpha\beta}_{1/2} \times V^{(1) \gamma\delta}_{1/2} \sim{}&
\epsilon_{\alpha\gamma}\epsilon_{\beta\delta} \big( w_8 I + w_9 V^{(1)}_0 + w_{10} T + w_{11} [A^{+i}{A^{+}}_i] + w_{12} [A^{-i}{A^{-}}_i] +{}\\ \notag&{}+ w_{13} [V^{(1)}_0V^{(1)}_0] + w_{14} V^{(2)}_0\big) +
\epsilon_{\beta\delta}\ell_{i,\alpha\gamma}\big(  w_{15} A^{+i} + w_{16} [A^{+i}V^{(1)}_0] +{}\\ \notag&{}+ w_{17} V^{(1)+i}_1 \big) +
\epsilon_{\alpha\gamma}\ell_{i,\beta\delta}\big( w_{18} A^{-i} + w_{19} [A^{-i}V^{(1)}_0] + w_{20} V^{(1)-i}_1 \big)+{}\\ \notag&{}+
\ell_{i,\alpha\gamma}\ell_{j,\beta\delta} w_{21}[A^{+i}A^{-j}]\ ,\\ \notag
V^{(1)}_0 \times V^{(2)}_0 \sim{}& w_{22}V^{(1)}_0 + w_{23}T + w_{24}[A^{+i}{A^{+}}_i] + w_{25}[A^{-i}{A^{-}}_i] + w_{26}[V^{(1)}_0V^{(1)}_0]+{}\\ &{}+
w_{27}V^{(2)}_0\ ,
\label{eq:ope3}
\end{align}
where the identity operator $I$ cannot appear in the last OPE because the two point function 
$\langle V^{(1)}_0(z) V^{(2)}_0(w)\rangle$ vanishes.
\smallskip

\noindent Similarly, the most general ansatz for the OPEs of total spin $\frac{7}{2}$ is
\begin{align}\notag
V^{(1)}_0\times V^{(1)\alpha\beta}_{3/2} \sim{}&
w_{28} G^{\alpha\beta} + w_{29} V^{(1)\alpha\beta}_{1/2} +
w_{30} V^{(1)\alpha\beta}_{3/2} + w_{31} V^{(2)\alpha\beta}_{1/2} + w_{32} [V^{(1)}_0 G^{\alpha\beta}] + {}\\\notag&{}+ w_{33} [V^{(1)}_0 V^{(1)\alpha\beta}_{1/2}]+
\rho_{i,\gamma\alpha} \big( w_{34} [A^{+i}V^{(1)\gamma\beta}_{1/2}] + w_{35} [A^{+i}G^{\gamma\beta}] \big)+ {}\\\notag&{}+
\rho_{i,\gamma\beta} \big( w_{36} [A^{-i}V^{(1)\alpha\gamma}_{1/2}]+
w_{37} [A^{-i}G^{\alpha\gamma}] \big)\ ,\\[4pt] \notag
V^{(1)}_0\times V^{(2)\alpha\beta}_{1/2} \sim{}& 
0\cdot  G^{\alpha\beta} + w_{38} V^{(1)\alpha\beta}_{1/2} + \cdots +
\rho_{i,\gamma\beta} \big( w_{45} [A^{-i}V^{(1)\alpha\gamma}_{1/2}] + w_{46}[A^{-i}G^{\alpha\gamma}] \big)\ ,\\[2pt] \notag
V^{(1)\alpha\beta}_{1/2}\times V^{(2)}_{0} \sim{}&
0\cdot  G^{\alpha\beta} +w_{47} V^{(1)\alpha\beta}_{1/2} + \cdots +
\rho_{i,\gamma\beta} \big( w_{54} [A^{-i}V^{(1)\alpha\gamma}_{1/2}]+ w_{55} [A^{-i}G^{\alpha\gamma}] \big)\ ,\\[4pt] \notag
V^{(1)\alpha\beta}_{1/2}\times V^{(1)+ i}_{1} \sim{}&
\rho^i_{\gamma\alpha}\big\{  w_{56} G^{\gamma\beta} +  w_{57} V^{(1)\gamma\beta}_{1/2}+
w_{58} V^{(1)\gamma\beta}_{3/2} + w_{59} V^{(2)\gamma\beta}_{1/2} + w_{60} [V^{(1)}_0G^{\gamma\beta}] + {}\\\notag&{}+ w_{61} [V^{(1)}_0 V^{(1)\gamma\beta}_{1/2}] +
\rho_{j,\delta\gamma}\big(w_{62} [A^{+j}V^{(1)\delta\beta}_{1/2}]+ w_{63} [A^{+j}G^{\delta\beta}] \big) + {}\\\notag&{}+
\rho_{j,\delta\beta} \big(w_{64} [A^{-j}V^{(1)\gamma\delta}_{1/2}]+ w_{65} [A^{-j}G^{\gamma\delta}] \big)\big\}+ {}\\\notag&{}+
 w_{66}[A^{+i}V^{(1)\alpha\beta}_{1/2}]+ w_{67} [A^{+i}G^{\alpha\beta}]\ ,\\[4pt] \notag
V^{(1)\alpha\beta}_{1/2}\times V^{(1)- i}_{1} \sim{}&
\rho^i_{\gamma\beta}\big\{  w_{68} G^{\alpha\gamma} + w_{69} V^{(1)\alpha\gamma}_{1/2}+
w_{70} V^{(1)\alpha\gamma}_{3/2}+w_{71} V^{(2)\alpha\gamma}_{1/2} + w_{72} [V^{(1)}_0G^{\alpha\gamma}]+ {}\\\notag&{}+
w_{73} [V^{(1)}_0 V^{(1)\alpha\gamma}_{1/2}] + \rho_{j,\delta\gamma}\big( w_{74} [A^{-j}V^{(1)\alpha\delta}_{1/2}]+ w_{75} [A^{-j}G^{\alpha\delta}] \big) +  {}\\\notag&{}+
\rho_{j,\delta\alpha} \big( w_{76} [A^{+j}V^{(1)\delta\gamma}_{1/2}]+w_{77} [A^{+j}G^{\delta\gamma}] \big)\big\}+{}\\
&{}+w_{78} [A^{-i}V^{(1)\alpha\beta}_{1/2}]+w_{79} [A^{-i}G^{\alpha\beta}]\ .
\label{eq:ope7h}
\end{align}
In order to explain the above notation we note that the general ansatz for the OPEs 
$V^{(1)}_0\times V^{(1)\alpha\beta}_{3/2}$, 
$V^{(1)}_0\times V^{(2)\alpha\beta}_{1/2}$ and $V^{(1)\alpha\beta}_{1/2}\times V^{(2)}_{0}$
all have the same form, except that the actual structure constants will in general be different; we have
therefore labelled the structure constants of the latter two OPEs using the same ordering as for the first.
We hope this compact notation does not lead to any confusion. We should also mention that 
the coefficient in front of $G^{\alpha\beta}$ in the OPE $V^{(1)}_0\times V^{(2)\alpha\beta}_{1/2}$ must vanish 
because the two operators belong to different superprimary multiplets and hence cannot generate the 
superconformal family of the identity. The same remark applies to the OPE $V^{(1)\alpha\beta}_{1/2}\times V^{(2)}_{0}$.
\smallskip

\noindent The general ansatz for the OPEs of total spin $4$ is given in Appendix~\ref{sec:A}.

\subsection{Jacobi Identities}\label{sec:jac}

Next we want to determine the actual structure constants, using the requirement that the ${\cal W}$ algebra
must have associative OPEs, i.e., $(A(z) B(w)) C(v)=A(z) (B(w) C(v))$. Using usual contour deformation
arguments, see e.g.~\cite{Thielemans:1994er}, this amounts to the condition that for all 
triplets  $A, B, C$ of $\mathcal{W}$ algebra generators we have the identity
\begin{equation}
[A[BC]_p]_q -(-1)^{|A||B|} [B[AC]_q]_p = \sum^\infty_{l=1} \binom{q-1}{l-1}[[AB]_lC]_{p+q-l}\ ,\quad p,q>0\ ,
\label{eq:def_Jacobi}
\end{equation}
where $[AB]_p$ is the operator that multiplies the $p$-th order pole in the OPE of $A$ with $B$, etc.
This condition is believed to be equivalent to the requirement that the corresponding Jacobi identities
are satisfied, and we shall denote the set of equations~\eqref{eq:def_Jacobi} by $A\times B\times C$.
To compute these identities we use the  packages \texttt{OPEdefs} and \texttt{OPEconf} of 
Thielemans, see \cite{Thielemans:1994er, Thielemans:1991uw}.
\smallskip

We shall proceed level by level. First we solve all the Jacobi identities that 
can be computed with the OPEs of sec.~\ref{sec:ope}. The first two OPEs \eqref{eq:ope2} allow
one to analyse the Jacobi identities
\begin{align*}
&T\times  V^{(1)}_0\times  V^{(1)}_0\ ,\quad A^{\pm i}\times V^{(1)}_0\times  V^{(1)}_0\ ,\quad 
V^{(1)}_0\times V^{(1)}_0\times  V^{(1)}_0\ ,\quad G^{\alpha\beta}\times V^{(1)}_0\times  V^{(1)}_0\ ,\\
&T\times V^{(1)}_0\times  V^{(1)\alpha\beta}_{1/2}\ ,\quad A^{\pm i}\times V^{(1)}_0\times  V^{(1)\alpha\beta}_{1/2}\ ,
\quad  V^{(1)}_0\times V^{(1)}_0\times  V^{(1)\alpha\beta}_{1/2}\ . 
\end{align*}
It turns out that all of these are trivially satisfied.
At one level higher, i.e.\ with the OPEs \eqref{eq:ope3}, one can compute the next group of Jacobi identities
\begin{align*}
&T\times V^{(1)}_0 \times V^{(1)\pm i}_1 \ ,\quad A^{\pm i}\times V^{(1)}_0 \times V^{(1)\pm j}_1 \ ,\quad 
V^{(1)}_0\times V^{(1)}_0 \times V^{(1)\pm j}_1 \ ,\\
&T\times V^{(1)\alpha\beta}_{1/2} \times V^{(1)\gamma\delta}_{1/2} \ ,\quad 
A^{\pm i}\times V^{(1)\alpha\beta}_{1/2} \times V^{(1)\gamma\delta}_{1/2} \ ,\quad 
V^{(1)}_0\times V^{(1)\alpha\beta}_{1/2} \times V^{(1)\gamma\delta}_{1/2} \ ,\\
&T\times V^{(1)}_0 \times V^{(2)}_0 \ ,\quad A^{\pm i}\times V^{(1)}_0 \times V^{(2)}_0 \ ,\quad 
V^{(1)}_0\times V^{(1)}_0 \times V^{(2)}_0 \ ,\quad 
G^{\alpha\beta}\times V^{(1)}_0 \times V^{(1)\gamma\delta}_{1/2}\ .
\end{align*}
These are satisfied provided the only non-zero structure constants in the OPEs \eqref{eq:ope3} are
\begin{align}
\notag
 n_{1}&=-\frac{2 k_- k_+}{2+k_-+k_+} \ ,&
 w_{1}&=1 \ ,&
 w_{2}&=-\frac{8 k_-}{2+k_-+k_+} \ ,\\ \notag
 w_{5}&=-\frac{8 k_+}{2+k_-+k_+} \ ,&
 w_{8}&=-\frac{8 k_- k_+}{2+k_-+k_+} \ ,&
 w_{10}&=-4 \ ,\\\notag
 w_{11}&=\frac{2}{2+k_-+k_+} \ ,&
 w_{12}&=\frac{2}{2+k_-+k_+} \ ,&
 w_{15}&=\frac{8 k_-}{2+k_-+k_+} \ ,\\ 
 w_{18}&=\frac{8 k_+}{2+k_-+k_+} \ ,&
 w_{21}&=-\frac{8}{2+k_-+k_+} \ , & 
 &
\label{eq:jac1}
\end{align}
where we have chosen to normalise $V^{(1)}_0$, and consequently all the other fields in the supermultiplet $V^{(1)}$, 
by fixing $w_1=1$. We remark that the only structure constant that is at this level not fixed is 
$w_{22}$. In fact, $w_{22}$ cannot  be determined in this manner because it reflects the 
freedom of redefining $V^{(2)}_0$ by a multiple of  $[V^{(1)}_0 V^{(1)}_0]+\cdots$, see 
eq.~\eqref{eq:n4comp}. We shall therefore, in the following, use this freedom to set 
\begin{equation}
w_{22}=0\ .
\label{eq:red1}
\end{equation}
Note that it follows from the structure of the OPEs~\eqref{eq:ope3} and the form of the 
structure constants~\eqref{eq:jac1} that no simple operator of spin $2$ appears on the r.h.s.\ of these OPEs.
As a consequence, we can also already now compute the special Jacobi identity
\begin{equation}
V^{(1)\alpha\beta}_{1/2}\times V^{(1)\gamma\delta}_{1/2}\times V^{(1)\mu\nu}_{1/2}\ .
\label{eq:jacsp1}
\end{equation}
However, as it turns out, this identity is automatically satisfied.

Next we turn to the Jacobi identities that can be computed with the OPEs~\eqref{eq:ope7h}.
In order to proceed efficiently, we first impose for all OPEs  $A\times B$ the Jacobi identity
$T\times A\times B$, i.e., we ensure that the conformal symmetry is respected. Then it follows from 
eq.~\eqref{eq:def_Jacobi} that in order to compute a Jacobi identity for a  triplet of generators $A, B, C$ for which 
the spins sum up to $s$, it is sufficient to know the OPEs between all pairs of generators for which the spins sum up to 
$s-1$. Thus, with the OPEs~\eqref{eq:ope7h} (as well as the OPEs from above), we can compute the Jacobi identities 
for all triplets of generators for which the spins sum up to $\frac{9}{2}$. 
Solving these identities we find that the non-zero structure constants in the OPEs~\eqref{eq:ope7h} must equal
\begin{align}\notag
 w_{28}&=-\frac{16 \left(-k_-+k_+\right)}{3 \left(2+k_-+k_+\right)} \ ,\\\notag
 w_{30}&=w_{70} \ ,\\\notag
 w_{31}&=1 \ ,\\\notag
 w_{32}&=-4 \left(-k_-+k_+\right) \left(5+4 k_-+4 k_++2 k_- k_+\right)K w_{70}
\ ,\\\notag
 w_{33}&=-4 \left(-k_-+k_+\right) \left(5+4 k_-+4 k_++2 k_- k_+\right)K
\ ,\\\notag
 w_{34}&=-\frac{8 \left(2+k_-+2 k_+\right) \left(-2-k_--k_++2 k_- k_++2 k_-^2 k_+\right) K w_{70}}{2+k_-+k_+ }
\ ,\\\notag
 w_{35}&=\frac{8 \left(2+k_-+2 k_+\right) \left(-2-k_--k_++2 k_- k_++2 k_-^2 k_+\right)K}{2+k_-+k_+}
\ ,\\\notag
 w_{36}&=\frac{8 \left(2+2 k_-+k_+\right) \left(-2-k_--k_++2 k_- k_++2 k_- k_+^2\right)K w_{70}}{2+k_-+k_+ }
\ ,\\\notag
 w_{37}&=-\frac{8 \left(2+2 k_-+k_+\right) \left(-2-k_--k_++2 k_- k_++2 k_- k_+^2\right)K}{2+k_-+k_+ }
\ ,\\\notag
 w_{39}&=-1-w_{70}^2 \ ,\\\notag
 w_{40}&=-w_{70} \ ,\\\notag
 w_{41}&=4 \left(-k_-+k_+\right) \left(5+4 k_-+4 k_++2 k_- k_+\right) \left(1+w_{70}^2\right)K
\ ,\\\notag
 w_{43}&=\frac{8 \left(2+k_-+2 k_+\right) \left(-2-k_--k_++2 k_- k_++2 k_-^2 k_+\right) \left(1+w_{70}^2\right)K}{2+k_-+k_+ }
 \ ,\\\notag
 w_{45}&=-\frac{8 \left(2+2 k_-+k_+\right) \left(-2-k_--k_++2 k_- k_++2 k_- k_+^2\right) \left(1+w_{70}^2\right)K}{2+k_-+k_+}
\ ,\\\notag
 w_{48}&=1+w_{70}^2 \ ,\\\notag
 w_{49}&=w_{70} \ ,\\\notag
 w_{50}&=-4 \left(-k_-+k_+\right) \left(5+4 k_-+4 k_++2 k_- k_+\right) \left(1+w_{70}^2\right)K
 \ ,\\\notag
 w_{52}&=-\frac{8 \left(2+k_-+2 k_+\right) \left(-2-k_--k_++2 k_- k_++2 k_-^2 k_+\right) \left(1+w_{70}^2\right)K}{2+k_-+k_+ }
 \ ,\\\notag
 w_{54}&=\frac{8 \left(2+2 k_-+k_+\right) \left(-2-k_--k_++2 k_- k_++2 k_- k_+^2\right) \left(1+w_{70}^2\right)K}{2+k_-+k_+}
 \ ,\\\notag
 w_{56}&=\frac{4 \left(1+k_-+2 k_+\right)}{2+k_-+k_+} \ ,\\\notag
 w_{58}&=-w_{70} \ ,\\\notag
 w_{59}&=-1 \ ,\\\notag
 w_{60}&=4 \left(-k_-+k_+\right) \left(5+4 k_-+4 k_++2 k_- k_+\right) w_{70}K
 \ ,\\\notag
 w_{61}&=4 \left(-k_-+k_+\right) \left(5+4 k_-+4 k_++2 k_- k_+\right)K
\ ,\\\notag
 w_{62}&=\frac{8 \left(2+k_-+2 k_+\right) \left(-2-k_--k_++2 k_- k_++2 k_-^2 k_+\right) K w_{70}}{2+k_-+k_+}
 \ ,\\\notag
 w_{63}&=-\frac{8 \left(2+k_-+2 k_+\right) \left(-2-k_--k_++2 k_- k_++2 k_-^2 k_+\right)K}{2+k_-+k_+}
 \ ,\\\notag
 w_{64}&=-\frac{8 \left(2+2 k_-+k_+\right) \left(-2-k_--k_++2 k_- k_++2 k_- k_+^2\right) Kw_{70}}{2+k_-+k_+ }
 \ ,\\\notag
 w_{65}&=\frac{8 k_- \left(-1+k_+\right) \left(1+k_+\right) \left(2+k_-+2 k_+\right)K}{2+k_-+k_+ }
 \ ,\\\notag
 w_{67}&=\frac{4}{2+k_-+k_+} \ ,\\\notag
 w_{68}&=\frac{4 \left(1+2 k_-+k_+\right)}{2+k_-+k_+} \ ,\\\notag
 w_{70}& \ ,\\\notag
 w_{71}&=1 \ ,\\\notag
 w_{72}&=-4 \left(-k_-+k_+\right) \left(5+4 k_-+4 k_++2 k_- k_+\right) Kw_{70}
 \ ,\\\notag
 w_{73}&=-4 \left(-k_-+k_+\right) \left(5+4 k_-+4 k_++2 k_- k_+\right)K
 \ ,\\\notag
 w_{74}&=\frac{8 \left(2+2 k_-+k_+\right) \left(-2-k_--k_++2 k_- k_++2 k_- k_+^2\right)K w_{70}}{2+k_-+k_+}
\ ,\\\notag
 w_{75}&=-\frac{8 \left(2+2 k_-+k_+\right) \left(-2-k_--k_++2 k_- k_++2 k_- k_+^2\right)K}{2+k_-+k_+ }
\ ,\\\notag
 w_{76}&=-\frac{8 \left(2+k_-+2 k_+\right) \left(-2-k_--k_++2 k_- k_++2 k_-^2 k_+\right) Kw_{70}}{2+k_-+k_+ }
 \ ,\\\notag
 w_{77}&=\frac{8 \left(-1+k_-\right) \left(1+k_-\right) k_+ \left(2+2 k_-+k_+\right)K}{2+k_-+k_+ }
 \ ,\\
 w_{79}&=\frac{4}{2+k_-+k_+}\ ,
\label{eq:jac2}
\end{align}
where $K$ is a shorthand notation for the frequently occurring expression
\begin{equation}
K = \frac{1}{-4-4 k_--k_-^2-4 k_++3 k_- k_++4 k_-^2 k_+-k_+^2+4 k_- k_+^2+3 k_-^2 k_+^2}\ ,
\label{eq:K}
\end{equation}
and we have chosen to normalize  $V^{(2)}$ by fixing $w_{31}=1$.
Notice that  the structure constants in the OPEs~\eqref{eq:ope7h} are uniquely determined by 
$k_\pm$ and $w_{70}$; there are no field redefinitions that render the structure constant $w_{70}$ redundant so, 
in principle it can either get fixed by the higher Jacobi identities or, if it does not, describe a genuine
parameter of the $\mathcal{W}_\infty$ algebra.

With the next set of OPEs~\eqref{eq:ope4} we can compute the Jacobi identities $A\times B\times C$ for all 
triplets of generators for which the spins sum up to $5$.
Solving these identities we find in particular that 
$$w_{129}=w_{161}=0\ ,$$
which means that  no $\mathcal{W}_\infty$ algebra generator of spin $3$ can appear in the singular part of the 
OPEs $V^{(2)}_0\times V^{(1)\pm i}_1$ 
and, obviously, also in $V^{(1)+ i}_1\times V^{(1)- j}_1$ and  $V^{(2)}_0\times V^{(2)}_0$.
For this reason,  the OPEs~\eqref{eq:ope4} are also sufficient to compute the special Jacobi identities
\begin{equation}
V^{(2)}_0\times V^{(2)}_0\times V^{(2)}_0\ ,\quad 
V^{(2)}_0\times V^{(2)}_0\times V^{(1)\pm i}_1\ ,\quad
V^{(2)}_0\times V^{(1)+i}_1\times V^{(1)-j}_1\ .
\label{eq:special2}
\end{equation}
Solving in addition these identities we find, first of all that
\begin{equation}
w_{70}=0\ ,
\label{eq:w70}
\end{equation}
and, secondly, that the following structure constants remain undetermined
\begin{equation}
w_{80}\ ,\quad w_{81}\ ,\quad w_{85}\ ,\quad w_{89}\ ,\quad w_{90}\ ,
\label{eq:param}
\end{equation}
while  all the other structure constants in the OPEs~\eqref{eq:ope4} are uniquely fixed in terms of these and $k_\pm$;
the explicit expressions for those structure constants that are non-zero are given in appendix~\ref{app:B1}.

Let us now  try to understand the meaning of the free parameters in eq.~\eqref{eq:param}.
Firstly, just like $w_{22}$, the structure constants $w_{80}$ and $w_{81}$ are redundant because they can be set
to any value by absorbing into $V^{(3)}_0$ a linear combination of the two composite $\mathcal{N}=4$
superprimary fields at spin $3$, see the discussion following eq.~\eqref{eq:n4comp}; we shall fix this redefinition 
freedom of $V^{(3)}_0$ by setting 
\begin{equation}
w_{80} = w_{81} = 0\ .
\label{eq:red2}
\end{equation}
Secondly, we note that there is a similarity between $w_{70}$ and $w_{85}$, $w_{89}$, $w_{90}$ --- they all appear in front 
of operators that violate the parity `symmetry' of the OPEs
\begin{equation}
V^{(s)}\mapsto (-1)^sV^{(s)}\ ,
\label{eq:parity}
\end{equation}
that is a natural symmetry of the underlying higher spin algebra $\mathfrak{shs}_2[\lambda]$. In fact, 
a careful inspection of the structure constants (\ref{eq:jac1}), (\ref{eq:jac2}), and (\ref{eq:w70}) shows that 
$w_{85}$, $w_{89}$, $w_{90}$ are the \emph{only} structure constants that violate this symmetry.
We have gone one level higher with the ansatz for the OPEs and  verified that, in perfect analogy 
with what happened to $w_{70}$, these parity violating structure constants are required to vanish by the next set of Jacobi identities
 for which the spins sum up to $\frac{9}{2}$
\begin{equation}
w_{85}=w_{89}=w_{90}=0\ .
\label{eq:conj}
\end{equation}
We interpret this fact as evidence for a mechanism by which the consistency of the $\mathcal{W}_\infty$ algebra 
imposes (dynamically) the parity  symmetry~\eqref{eq:parity} on all the OPEs. 
Furthermore, if we assume that the parity symmetry~\eqref{eq:parity} holds generally,
then all the structure constants could again be determined uniquely in terms of 
$k_\pm$, modulo the redefinition freedom of the generators. 
We take this as a strong indication that the most general $\mathcal{N}=4$ $\mathcal{W}_\infty$ algebra with the 
above field content does not have any other parameters except for $k_\pm$.


\subsection{Truncations}


Given the higher spin / CFT duality of \cite{Gaberdiel:2013vva} we expect the large $\mathcal{N}=4$ 
$\mathcal{W}_\infty$ algebra
to exhibit two kinds of truncations. First, for suitable values of $\mu$ (namely $\mu\in\mathbb{Z} \backslash \{0,1\}$), the underlying
higher spin algebra $\mathfrak{shs}_2[\mu]$ can be truncated to a finite dimensional Lie algebra, and one may 
therefore expect that this will also be reflected in the corresponding $\mathcal{W}_\infty$ algebra. Second, the dual
Wolf space cosets should be finitely generated, and thus we should expect the ${\cal W}_\infty$ algebra
to truncate for positive integer values of $k_\pm$. Unlike the situation with less supersymmetry
\cite{triality,Candu:2012tr}, these two truncation phenomena seem to be of different nature (see also the 
discussion in \cite{Gaberdiel:2013vva}), and we shall therefore study them separately.

\subsubsection{The Higher Spin Truncation}

As already explained in  \cite{Gaberdiel:2013vva}, if we set $\mu=-N$ or $\mu=N+1$ with $N\in\mathbb{N}$, then 
the higher spin algebra $\mathfrak{shs}_2[\mu]$ can be truncated to an algebra that is generated by $D(2,1;\alpha)$, 
the first $N-1$ supermultiplets $V^{(s)}$ with $s=1,\dots,N-1$, as well as `half' of the $N$-th supermultiplet
\begin{equation}
\hat{V}^{(N)\pm }= \{\,V^{(N)}_0 \ ,\ V^{(N)\alpha\beta}_{1/2}\ ,\  V^{(N)\pm i}_1\,\}\ ,
\label{eq:halfV}
\end{equation}
where the plus case arises for $\mu=-N$ and the minus case for $\mu=N+1$.

Let us concentrate in the following on the case $\mu=N+1$ for which the minus truncation of (\ref{eq:halfV}) 
arises; the other case works similarly. In order for the multiplet to truncate in the actual ${\cal W}_\infty$
algebra the missing states, i.e., the states that would be there in $V^{(N)}$ but are absent in $V^{(N)-}$, 
must actually be null; thus the higher spin analysis predicts null-vectors which turn out to be of the form
(see also \cite{Gaberdiel:2014yla})
\begin{align}
s&=N+1: &&V^{(N)-i}_1 + \kappa [A^{-i}V^{(N)}_0]\ , \notag \\
s&=N+\tfrac{3}{2}: && V^{(N)\alpha\beta}_{3/2}+ \kappa  [G^{\alpha\beta}V^{(N)}_0]  -2 \kappa\rho_{i,\gamma\alpha}[A^{+i}V^{(N)\gamma\beta}_{1/2}]\ , \notag\\
s&=N+2: && V^{(N)}_2 - 4\kappa [TV^{(N)}_0] -\kappa \epsilon_{\alpha\gamma}\epsilon_{\beta\delta} [G^{\alpha\beta}V^{(N)\gamma\delta}_{1/2}] + 
\tfrac{1}{2}\kappa^2 [[{A^+}_i A^{+i}]V^{(N)}_0]+ {} \notag \\ &&& \qquad {}+ 
\tfrac{1}{2}\kappa^2  [[{A^-}_i A^{-i}]V^{(N)}_0]-\kappa  [{A^{+}}_i V^{(N)+i}_1] \ ,
\label{eq:halfVnull}
\end{align}
where  $\kappa = -4 N/k_-$. These solutions appear for  $k_-$ given by\footnote{There is a 
second solution $k_-=1$ to which we will return below. This fact was also noticed in \cite{Gaberdiel:2014yla}.}
\begin{equation}
k_- = -\frac{N (2 + k_+)}{1 + N}\ .
\label{eq:kpkmrel}
\end{equation}
Here the $\kappa$-dependent terms are required to make the states
in eq.~\eqref{eq:halfVnull} primary with respect to the stress-energy tensor and the current fields. In the 
`t~Hooft limit, $k_\pm \rightarrow \infty$ with the ratio $\alpha = k_-/k_+$ kept fixed, 
$\alpha\to -N/(1+N)$ which corresponds precisely to $\mu=N+1$. In this limit the constant
$\kappa$ vanishes and we recover the truncation exhibited in  \cite{Gaberdiel:2013vva}.

We have checked that these vectors are singular with respect to the (non-linear) large ${\cal N}=4$ superconformal
algebra, but we expect that they actually lie in an ideal of the full ${\cal W}_\infty$ algebra (that can be consistently
quotiened out). Given our detailed understanding of the latter, we can check this at least for $N=1$ and $N=2$. 
To illustrate these checks, consider first the case $N=1$. It follows from \eqref{eq:ope3}  that 
\begin{equation}
V^{(1)}_0 \times \big(V^{(1)-i}_1 + \kappa [A^{-i}V^{(1)}_0]\big) = (w_5 - \kappa n_1)A^{-i} \ , 
\label{eq:tr1}
\end{equation}
where we have used that $[A^{-i}V^{(1)}_0]=(A^{-i}V^{(1)}_0)$, as well as  
$w_6=w_7=0$, see eq.~\eqref{eq:jac1}. Since the left-hand-side should lie in the ideal (but $A^{-i}$ does not)
consistency requires that the prefactor vanishes, $w_5-\kappa n_1=0$; this turns out to be true, using 
eq.~\eqref{eq:jac1} for $N=1$. 

A somewhat more trivial test is that in the OPE $V^{(1)}_0\times V^{(2)}_0$ the coefficient $w_{22}$ 
of $V^{(1)}_0$ vanishes; this is automatically the case
for our definition of $V^{(2)}_0$, see the discussion around eq.~(\ref{eq:red1}). Less trivially, 
in the OPEs 
\be
V^{(1)}_0  \times V^{(2)}_{1/2} \ , \qquad \hbox{and} \qquad V^{(1)}_{1/2}  \times V^{(2)}_{0}
\ee
in eq.~\eqref{eq:ope7h}, the right-hand-side is indeed null because 
the coefficients of the terms that do not break parity, i.e., $V^{(1)}_{1/2}$ and  $[A^{-i}V^{(1)}_{1/2}]$ vanish, and 
$V^{(1)}_{3/2}$ enters only in the combination~\eqref{eq:halfVnull}, i.e.\ 
$w_{41}/w_{39}= w_{50}/w_{48}=\kappa$ and $w_{43}/w_{39}=w_{52}/w_{48}=-2\kappa$.
The fact that the whole multiplet $V^{(2)}$ is null also follows from the vanishing of the  central term 
$n_2$ in the OPEs $V^{(2)}_0\times V^{(2)}_0$, see appendix~\ref{app:B1}.
\smallskip

The analysis for $N=2$ is similar, except for one interesting subtlety. The only OPE on which 
we can test this truncation is $V^{(1)}_0\times V^{(2)-i}_1$, for which we find 
\begin{equation}
V^{(1)}_0 \times \big(V^{(2)-i}_1 + \kappa [A^{-i}V^{(2)}_0]\big) = w_{143} V^{(1)-i}_1 + (w_{144} +\kappa w_{22})[A^{-i}V^{(1)}_0]+ w_{152}\epsilon_{\alpha\gamma}r^i_{\beta\delta}[G^{\alpha\beta}V^{(1)\gamma\delta}_{1/2}]\ .
\label{eq:trouble1}
\end{equation}
The right hand side does not depend on $\kappa$ because  $V^{(1)}_0$ has a regular OPE with $V^{(2)}_0$,
and, on the face of it, it does not vanish. This is a consequence of the fact that the actual null-vector of the 
full ${\cal W}_\infty$ algebra requires a specific choice for $V^{(2)}_0$, which in the above conventions corresponds
not to $w_{22}=0$ (see eq.~(\ref{eq:red1})), but rather to  
\begin{equation}
w_{22} = -\frac{32 (k_+-1) (1 + 2 k_+)}{3 k_+ (2 + k_+)} \ . 
\end{equation}
With this choice of $V^{(2)}_0$ and setting $k_-$ to equal eq.~\eqref{eq:kpkmrel} with $N=2$, 
the right-hand-side of (\ref{eq:trouble1}) is indeed zero, i.e., 
$w_{143}=w_{144}+\kappa w_{22}=w_{152}=0$.

\subsubsection{The Coset Truncation}

Recall that the Wolf space coset algebra (written in bosonic form) 
\begin{equation}
\frac{\mathfrak{su}(N+2)_k\oplus \mathfrak{so}(4N)_1}{\mathfrak{su}(N)_{k+2}\oplus\mathfrak{u}(1)}
\label{eq:coset}
\end{equation}
has a non-linear large $\mathcal{N}=4$ superconformal symmetry with $k_+=k$ and $k_-=N$. 
For $k$ and $N$ large, the higher spin content of the above coset algebra agrees with the
large $\mathcal{N}=4$ $\mathcal{W}_\infty$ algebra  for spins sufficiently small compared to $k$ and $N$, 
see \cite{Gaberdiel:2013vva} for a simple higher
spin counting argument or \cite{Candu:2013fta} for a more involved
proof based on characters. Furthermore, it was confirmed in \cite{Gaberdiel:2014yla}, that
the asymptotic symmetry algebras match. 
It is therefore very natural to expect that the $\mathcal{N}=4$ $\mathcal{W}_\infty$ algebra truncates to the 
above  coset algebra at positive integer levels $k_\pm$.
The case with $N=3$ was also discussed in \cite{Ahn:2013oya}.

The first hint on the form of the coset truncation can be obtained by comparing  the vacuum character of the coset algebra~\eqref{eq:coset} to
the $\mathcal{W}_\infty$ algebra.
The first deviation can be computed with the help of eq.~(3.24)  of \cite{Candu:2013fta} and the  $\mathfrak{su}(N)$ 
modification rules of \cite{King:1971rs}, and for large enough $k$ one finds
\begin{equation}
\chi_{\text{coset}} = \chi_{\infty} - q^{N+1} \sum_{l_-=0}^{N+1}\ch_{l_-}(z_-) + \mathcal{O}(q^{N+\frac{3}{2}})\ .
\label{eq:nullpred}
\end{equation}
For example, for $k_-=N=1$ these null vectors appear at conformal dimension $2$ and are given  by 
\begin{align}
(l_+,l_-)=(0,2):&\qquad [A^{-i}A^{-j}]+[A^{-j}A^{-i}]-\tfrac{2}{3}\eta^{ij}[{A^{-}}_lA^{-l}]\ ,\notag\\
(l_+,l_-)=(0,1):&\qquad V^{(1)-i}_1 - 4 [A^{-i}V^{(1)}_0]\ ,\notag\\
(l_+,l_-)=(0,0):&\qquad  V^{(2)}_0\ ,
\label{eq:nullkm1}
\end{align}
where the first vector corresponds to the lowest affine null vector of the $\mathfrak{su}(2)_1$ vacuum representation, 
while the fact that $V^{(2)}_0$ is null follows from the vanishing of its 2-point function, i.e., $n_2$ vanishes 
(without inducing poles in any other structure constants).
The existence of the second null vector, explained by the fact that the representation of $\mathfrak{su}(2)_{k_-}$ 
with spin $l_-=1$ is not integrable at $k_-=1$, implies that $V^{(1)}$
truncates to a short representation $\hat V^{(1)}$ (and, in fact, all supermultiplets $V^{(s)}$ truncate this way).
The null vectors~\eqref{eq:nullkm1} and the ideal generated by them suggest that only the generators of the 
superconformal algebra and $\hat V^{(1)}$ survive the truncation, although, in contradistinction with the situation
in the previous subsection, the remaining generators must satisfy infinitely many additional constraints to 
account for the affine $\mathfrak{su}(2)_1$ null vectors.
\smallskip

For $k_-=2$ the first set of null vectors that are predicted by eq.~\eqref{eq:nullpred} appear at conformal dimension 
$3$, and 
they correspond to the lowest null vector of the $\mathfrak{su}(2)_2$ vacuum representation, 
which has spin $(l_+,l_-)=(0,3)$,  the lowest null vector of the  $\mathfrak{su}(2)_2$ representation 
generated by the affine primary $V^{(2)-i}_1 - 4 [A^{-i}V^{(2)}_0]$,  which has spin $(l_+,l_-)=(0,2)$, 
the unique superprimary with spin $(l_+,l_-)=(0,1)$ at conformal dimension $3$
\begin{eqnarray}
&& (V^{(2)-i}_1 -4 [A^{-i}V^{(2)}_0]) + 
\tfrac{4(k_++1)(10+11k_+)}{19 k_+^2 +18 k_+ -16}\left([V^{(1)}_0 (V^{(1)-i}_1-
2[A^{-i}V^{(1)}_0])]\right)+{} \notag\\
&&\qquad {}+\tfrac{8 (k_++6) (2 k_+^2-5)\epsilon_{\alpha\gamma}r^i_{\beta\delta}}{(k_++4) (19 k_+^2 +18 k_+ -16)} \left(
[V^{(1)\alpha\beta}_{1/2}V^{(1)\gamma\delta}_{1/2}]-
\tfrac{k_+^2+3k_+-1}{2k_+^2-5}[G^{\alpha\beta}G^{\gamma\delta}]
\right)+{} \notag\\
&&\qquad {}+\tfrac{32 ( k_+-4) (k_++1) (k_++6) }{(k_++4) (19 k_+^2 +18 k_+ -16)}[TA^{-i}]
-\tfrac{16 (k_+-4) k_+ ( k_++1) (k_++6)}{
 5 (k_++4)^2 (19 k_+^2 +18 k_+ -16)}{f^i}_{jl}[A^{-j}A^{-l}]_{-1}-{} \notag\\
&&\qquad {}-\tfrac{16 (k_+-4) (k_++1) (k_++5) (k_++6)}{
 5 (k_++4)^2 19 k_+^2 +18 k_+ -16)}\left([[{A^{-}}_jA^{-j}]A^{-i}]+
\tfrac{5}{(k_++5)}[[{A^{+}}_jA^{+j}]A^{-i}]\right) \ , 
\label{eq:null01}
\end{eqnarray}
as well as $V^{(3)}_0$ (with spin $(l_+,l_-)=(0,0)$). 
A non-trivial check of the  fact that the latter two superprimaries are null is that their OPEs with $V^{(1)}_0$ vanish
indeed. Again, it is tempting to believe that only the  superconformal algebra and the first two 
supermultiplets $V^{(1)}$ and $V^{(2)}$ survive the truncation, but it is clear from the affine representation
theory that they must satisfy infinitely many additional  constraints.
\smallskip

In general, we expect that for arbitrary integer values of $k_\pm$ the coset algebra is generated by 
$A^{\pm i}$, $T$, $G^{\alpha\beta}$ as well as the first $\min(k_-,k_+)$ supermultiplets. Furthermore, 
there are infinitely many additional constraints of spin $s\geq \min(k_-,k_+)$, accounting for the 
various $\mathfrak{su}(2)_{k_\pm}$ null vectors.

\section{Linear Large $\mathcal{N}=4$ $\mathcal{W}_\infty$ algebra}\label{sec:n4}

In the previous sections we have discussed the structure of the `non-linear' large ${\cal N}=4$
${\cal W}_\infty$ algebra that contains, in addition to the non-linear large ${\cal N}=4$ superconformal
algebra $\tilde{A}_\gamma$, muliplets of spin $s=1,2,\ldots$. The non-linear large ${\cal N}=4$ superconformal
algebra $\tilde{A}_\gamma$ can be obtained, upon quotienting out the free fermions and the $\mathfrak{u}(1)$ 
current \cite{goddard} from the linear $A_\gamma$ algebra, see 
\cite{Sevrin:1988ew,Schoutens:1988ig,Spindel:1988sr,VanProeyen:1989me,Sevrin:1989ce,goddard}
for some early literature on the subject. The same construction can also be applied to the `linear version'
of the full ${\cal W}_{\infty}$ algebra. One may therefore suspect that the structure of the 
`linear' ${\cal W}_{\infty}$ algebra will also be characterised just by the levels $k_\pm$ of the
two affine $\mathfrak{su}(2)$ algebras.\footnote{To avoid confusion we should stress that the full
`linear' ${\cal W}_{\infty}$ algebra is in fact also non-linear --- by the qualifier `linear' we only mean that 
it contains the linear large ${\cal N}=4$ superconformal algebra as a subalgebra (rather than the
non-linear $\tilde{A}_\gamma$ algebra). The fact that
this algebra cannot be completely linearised was already noticed, on the level of the dual asymptotic
symmetry algebra, in \cite{Gaberdiel:2014yla}.}

In order to confirm this we shall, in this section, repeat the above analysis for the linear case. Since
the techniques are largely the same, we shall be relatively brief.

\subsection{The Linear $\mc N=4$ Superconformal Algebra $A_{\gamma}$}\label{sec:l4}

The linear large ${\cal N}=4$ superconformal algebra $A_{\gamma}$ contains in addition to the energy momentum tensor, 
the current algebra
\be
\mk{su}(2)_{k_{+}}\oplus \mk{su}(2)_{k_{-}}\oplus\mk{u}(1)\ ,
\ee
as well as four supercharges $G^{a}$  and four free fermions $Q^a$ both of which  transform in the  
$(\frac{1}{2},\frac{1}{2})_{0}$ with respect to the above current algebra. We shall denote the 
$\mathfrak{u}(1)$ current by $U$, while the currents of $\su(2)\oplus\su(2)$ are 
$A^{\pm,i}$ with $i=1,2,3$. The central charge of the Virasoro algebra equals 
\be
c = \frac{6\,k_{+}\,k_{-}}{k_{+}+k_{-}}\ ,\quad \hbox{and} \quad \gamma=\frac{k_{-}}{k_{+}+k_{-}}\ ,
\qquad (\overline\gamma=1-\gamma)\ .
\ee
%
%
%
Apart from the standard $TT$ OPE, the additional OPEs defining $A_{\gamma}$ are
\ba
G^{a}(z)\,G^{b}(w) &\sim& \frac{2c}{3}\,\frac{\delta^{ab}}{(z-w)^{3}}-8\,\frac{\gamma\,\alpha_{ab}^{+,i}\,
A^{+,i}+\overline\gamma\,\alpha^{-,i}_{ab}\,A^{-,i}}{(z-w)^{2}}\nonumber \\
&& -4\,\frac{\gamma\,\alpha^{+,i}_{ab}\,
\partial A^{+,i}+\overline\gamma\,\alpha^{-}_{ab}\,\partial A^{-,i}}{z-w}+\frac{2\,\delta^{ab}\,T}{z-w}\ , \label{GGlin}
\\
A^{\pm,i}(z)\,A^{\pm,j}(w) &\sim& -\frac{k^{\pm}}{2}\,\frac{\delta^{ij}}{(z-w)^{2}}+\frac{\epsilon^{ijk}\,A^{\pm, k}}{z-w}\ ,\\
Q^{a}(z)\,Q^{b}(w) &\sim& -\frac{k^{+}+k^{-}}{2}\,\frac{\delta^{ab}}{z-w}\ ,\\
U(z)\,U(w) &\sim& -\frac{k^{+}+k^{-}}{2}\,\frac{1}{(z-w)^{2}}\ ,\\
A^{\pm, i}(z)\,G^{a}(w) &\sim& 
\mp\frac{2\,k^{\pm}}{k^{+}+k^{-}}\,\frac{\alpha^{\pm,i}_{ab}\,Q^{b}}{(z-w)^{2}}+\frac{\alpha^{\pm,i}_{ab}\,G^{b}}{z-w}\ , \\
A^{\pm, i}(z)\, Q^{a}(w) &\sim& \frac{\alpha^{\pm, i}_{ab}\,Q^{b}}{z-w}\ , \\
Q^{a}(z)\,G^{b}(w) &\sim& 2\,\frac{\alpha^{+,i}_{ab}\,A^{+,i}-\alpha^{-,i}_{ab}\,A^{-,i}}{z-w}
+\frac{\delta^{ab}\,U}{z-w}\ ,\\
Q^{a}(z)\,U(w) &\sim& 0\ , \\
U(z)\,G^{a}(w) &\sim& \frac{Q^{a}}{(z-w)^{2}} \ .
\ea
Here the matrices $\alpha^{\pm,i}_{ab}$ are the $\mk{so}(4)$ generators\footnote{In our
conventions $\epsilon_{123}=\epsilon^{123}=1$ and $\epsilon_{ab4}=0$.}
\be
\alpha^{\pm,i}_{ab} = \tfrac{1}{2}(\pm\delta^{i}_{a}\,\delta^{4}_{b}\mp\delta^{i}_{b}\,\delta^{4}_{a}+\epsilon_{iab})\ ,
\ee
obeying the (anti)-commutation relations
\be
[\alpha^{\pm,i},\alpha^{\pm,j}]=-\epsilon^{ijk}\,\alpha^{\pm,k}\ ,\qquad
[\alpha^{\pm,i},\alpha^{\mp,j}]=0\ ,\qquad
\{\alpha^{\pm,i},\alpha^{\pm,j}\}=-\tfrac{1}{2}\delta^{ij}\ .
\ee

\subsection{The General Linear Multiplet}
\label{sec:LinMult}

For the description of the linear ${\cal W}_\infty$ algebra we now need to add a linear 
$\mc N=4$ multiplet whose components close under the OPE with $A_{\gamma}$ 
\cite{Ivanov:1992rt,Ivanov:1989qs} (see also  \cite{Nagi:2004wb}). As before, we 
only need the special case of a  scalar multiplet, i.e., one whose lowest component is 
$\mk{su}(2)\oplus\mk{su}(2)$ invariant. The multiplet components can be labelled as in 
eq.~(\ref{eq:nonlinear-components}), except that we use a different convention
to label the $\mk{su}(2)\oplus\mk{su}(2)$ indices
\be\label{lincomp}
\begin{array}{c|ccccc}
{\rm component} & V^{(s)}_{0} & V^{(s), a}_{1/2} & V^{(s), \pm, i}_{1} & V^{(s), a}_{3/2} & V^{(s)}_{2}   \\ \hline 
\text{ conformal spin }    &   s & s+\frac{1}{2} & s+1 & s+\frac{3}{2} & s+2\\ 
\text{$\mathfrak{su}(2)\oplus\mathfrak{su}(2)$  spin } & (0,0) & (\frac{1}{2},\frac{1}{2}) & (1,0)\oplus(0,1) & (\frac{1}{2},\frac{1}{2}) & (0,0) 
\end{array}
\ee
The OPEs of the $A_\gamma$ fields with the various component fields are given in Appendix~\ref{app:linsup}.
\medskip

We can then proceed as in the analysis of the non-linear algebra. We make the most general ansatz for
the OPEs between the various higher spin fields, and then impose Jacobi identities, i.e., the associativity
of the OPEs, to determine the structure constants recursively. We have performed this analysis for the 
OPEs up to total spin $\frac{7}{2}$. We have again found that, apart from $k_\pm$, there are no free 
parameters --- except for those one would expect to be determined by imposing higher Jacobi identities.

\subsection{From the Linear to the Non-Linear Description}

According to \cite{goddard}, it is possible to decouple the free fermions $Q^{a}$, and the 
$\mathfrak{u}(1)$ field $U$ from
the linear ${\cal W}_{\infty}$ algebra by effectively performing a coset construction. On the level
of the linear superconformal $A_\gamma$ algebra, this amounts to redefining the stress-energy tensor, the supercharges
and the affine currents as 
\ba
\label{eq:Ttilde}
\widetilde T &=& T+\, \frac{1}{k_{+}+k_{-}}\,\big[-(Q^{c}\partial Q^{c})+ (UU)\big]\ , \\
\label{eq:Gtilde}
\widetilde G^{a} &=& G^{a}+\frac{2}{k_{+}+k_{-}}\,\Big[
(UQ^{a})-2\,\alpha^{+,i}_{ab}(A^{+,i}Q^{b})
+2\,\alpha^{-,i}_{ab}(A^{-,i}Q^{b})\nonumber \\
&&\qquad +\frac{2}{3}\frac{1}{k_{+}+k_{-}}\,\epsilon_{abcd}(Q^{b}Q^{c}Q^{d})
\Big]\ , \\
\label{eq:Atilde}
\widetilde{A}^{\pm,i} &=& A^{\pm,i}-\frac{1}{k_{+}+k_{-}}\,\alpha^{\pm,i}_{ab}(Q^{a}Q^{b})\ .
\ea
The modified fields then obey the non-linear $\widetilde{A}_{\gamma}$ algebra and 
have regular OPEs with the decoupling fields
\ba
U(z)\,\{\widetilde A^{\pm, i}(w), \ \widetilde G^{a}(w),\ \widetilde T(w)\}&\sim& 0\ , \\
Q^{a}(z)\,\{\widetilde A^{\pm, i}(w), \ \widetilde G^{b}(w),\ \widetilde T(w)\}&\sim& 0\ .
\ea
The redefined currents still satisfy an affine $\su(2)\oplus \su(2)$ algebra, but the levels are now shifted to 
\be\label{eq:levelrel}
\tilde k_{\pm} = k_{\pm}-1\ .
\ee
\medskip

We can similarly decouple the free fermions and the $\mathfrak{u}(1)$ field from the rest of the linear
${\cal W}_{\infty}$ algebra. For the lowest spin component there is nothing to be done, 
\be\label{eq:Vstilde}
\widetilde V^{(s)}_{0} = V^{(s)}_{0}\ ,
\ee
and the remaining  components can be obtained by repeatedly applying the supercurrents $\widetilde G^{a}$; this leads to 
\ba
\label{eq:V12}
\widetilde V^{(s),a}_{1/2} &=& V^{(s), a}_{1/2}\ , \\
\label{eq:V1}
\widetilde V^{(s), \pm, i}_{1} &=& V^{(s), \pm, i}_{1}\pm\frac{4}{k_{+}+k_{-}}\,\alpha^{\pm,i}_{ab}(Q^{a}\,V^{(s),b}_{1/2})\ , \\
\label{eq:V32}
\widetilde V^{(s),a}_{3/2} &=&V^{(s),a}_{3/2}+\frac{4}{k_{+}+k_{-}}\Big[
2\,s\,\,(\partial Q^{a}\,V^{(s)}_{0})
-(Q^{a}\,\partial V^{(s)}_{0})
-(U\,V^{(s),a}_{1/2}) \\
&&
-2\alpha^{+,i}_{ab}( A^{+,i}\,V^{(s),b}_{1/2})
+2\alpha^{-,i}_{ab}( A^{-,i}\,V^{(s),b}_{1/2})
-\alpha^{+,i}_{ab}(Q^{b}\,V^{(s),+,i}_{1})
-\alpha^{-,i}_{ab}(Q^{b}\,V^{(s),-,i}_{1})
\Big]\ ,\nonumber \\
\label{eq:V2}
\widetilde V_{2}^{(s)} &=& V_{2}^{(s)} +\frac{4}{k_{+}+k_{-}}\Big[
-(2s+1)\,(\partial Q^{a}\,V^{(s),a}_{1/2})
+(Q^{a}\,\partial V^{(s),a}_{1/2})\nonumber \\
&&
+2\,s\,(\partial U\,V_{0}^{(s)})
-2\,( U\,\partial V_{0}^{(s)})
\Big]\ .
\ea
By construction, these component fields then have regular OPEs with the free fermions and the $\mathfrak{u}(1)$
field, as one may also check directly,
\ba
U(z)\,\widetilde V^{(s)}(w) \sim 0 \ , \qquad 
Q^{a}(z)\,\widetilde V^{(s)}(w) \sim 0 \ .
\ea

\subsection{Comparison of the Structure Constants}

As a cross-check of our results we should now be able to reproduce the 
OPEs of the non-linear ${\cal W}_{\infty}$ algebra from those of the linear analysis. 
Up to the level to which we have determined the linear algebra\footnote{Since the linear
algebra contains more fields, it is harder to push the analysis to the same level as for 
the non-linear case.} we have performed this analysis, and we have found perfect agreement,
thus giving a highly non-trivial consistency check on our analysis. 
In order to illustrate the nature of the comparison, let us give two specific examples.
\smallskip

%
%

The simplest case is the fusion of  the first multiplet $V^{(1)}$ with itself. Up to the level considered below, 
only the conformal block of the identity appears, and the first few cases are explicitly 
\ba
\label{eq:lin-ope1}
V^{(1)}_{0}\,V^{(1)}_{0} &\sim& \frac{1}{z^{2}}\,\mc O_{0}+\frac{1}{z}\,\mc O_{1}\ , \\
\label{eq:lin-ope2}
V^{(1)}_{0}\,V^{(1), a}_{1/2} &\sim& \frac{1}{z^{2}}\,\mc O^{a}_{1/2}+\frac{1}{z}\,\mc O^{a}_{3/2}\ ,\\
\label{eq:lin-ope3}
V^{(1), a}_{1/2}\,V^{(1), b}_{1/2} &\sim& \frac{1}{z^{3}}\,\mc O^{ab}_{0}+\frac{1}{z^{2}}\,\mc O^{ab}_{1}
+\frac{1}{z}\,\mc O^{ab}_{2}\ ,\\
\label{eq:lin-ope4}
V^{(1)}_{0}\,V^{(1), \pm, i}_{1} &\sim& \frac{1}{z^{2}}\,\mc O^{\pm, i}_{1}+\frac{1}{z}\,\mc O^{\pm, i}_{2}\ ,
\ea
where $\mc O_{s}$ is an operator of dimension $s$ built with the components of $A_{\gamma}$.
The solution of the Jacobi identities predicts that the operators on the right hand side are 
\ba
\label{eq:110-solution}
\mc O_{0} &=& n_{1}\,\mb I\ ,  \\
\mc O_{1} &=& 0\ , \nonumber \\
\mc O_{1/2}^{a} &=& 0\ , \nonumber  \\
\mc O_{3/2}^{a} &=& z_{1}\,G^{a}+z_{2}\,(UQ^{a})+z_{3}\,\alpha^{+,i}_{ab}(A^{+,i}Q^{b})
+z_{4}\,\alpha^{-,i}_{ab}(A^{-,i}Q^{b})+z_{5}\,\epsilon_{abcd}(Q^{b}Q^{c}Q^{d})\ ,\nonumber \\
\mc O_{0}^{ab} &=& z_{6}\,\delta^{ab}\,\mb I\ , \nonumber\\
\mc O_{1}^{ab} &=& z_{7}\,\alpha^{-,i}_{ab}A^{-,i}+z_{8}\,\alpha^{+,i}_{ab}A^{+,i}+z_{9}\,(Q^{a}Q^{b})
+z_{10}\,\epsilon_{abcd}(Q^{c}Q^{d})\ ,\nonumber\\
\mc O_{2}^{ab} &=& z_{11}\,\delta^{ab}(A^{-,i}A^{-,i})+z_{12}\,\alpha^{-,i}_{ac}\alpha^{+,j}_{cb}\,(A^{-,i}A^{+,j})
+z_{13}\,\delta^{ab}(A^{+,i}A^{+,i})+z_{14}\,\alpha^{-,i}_{ab}\partial A^{-,i}\nonumber \\
&&+z_{15}\,\alpha^{-,i}_{ac}(A^{-,i}Q^{c}Q^{b})+z_{16}\,\alpha^{-,i}_{ac}\epsilon_{cbde}(A^{-,i}Q^{d}Q^{e})
+z_{17}\,\delta^{ab}\alpha^{-,i}_{cd}(A^{-,i}Q^{c}Q^{d})\nonumber\\
&&+z_{18}\,\alpha^{+,i}_{ac}(A^{+,i}Q^{c}Q^{d})+z_{19}\,\alpha^{+,i}_{ac}\epsilon_{cbde}(A^{+,i}Q^{d}Q^{e})
+z_{20}\,\delta^{ab}\alpha^{+,i}_{cd}(A^{+,i}Q^{c}Q^{d})\nonumber \\
&&+z_{21}\,\alpha^{+,i}_{ab}\partial A^{+,i}+z_{22}\,(Q^{a}\partial Q^{b})+z_{23}\,(\partial Q^{a}Q^{b})
+z_{24}\,\delta^{ab}(Q^{c}\partial Q^{c})\nonumber\\
&& + z_{25}\,\epsilon_{abcd}(Q^{c}\partial Q^{d})+z_{26}\,\delta^{ab}T+z_{27}\,\delta^{ab}(UU)\ ,\nonumber\\
\mc O^{+,i}_{1} &=& z_{28}\,A^{+,i}+z_{29}\,\alpha^{+,i}_{ab}(Q^{a}Q^{b})\ , \nonumber\\
\mc O^{+,i}_{2} &=& z_{30}\,\epsilon^{ijk}\alpha^{+,j}_{cd}(A^{+,k}Q^{c}Q^{d})
+z_{31}\,\alpha^{+,i}_{ab}(\partial Q^{a}Q^{b})+z_{32}\,\alpha^{+,i}_{ab}(Q^{a}G^{b})
+z_{33}\,\alpha^{+,i}_{ab}(UQ^{a}Q^{b})\ ,\nonumber\\
\mc O^{-,i}_{1} &=& z_{34}\,A^{-,i}+z_{35}\,\alpha^{-,i}_{ab}(Q^{a}Q^{b})\ , \nonumber\\
\mc O^{-,i}_{2} &=& z_{36}\,\epsilon^{ijk}\alpha^{-,j}_{cd}(A^{-,k}Q^{c}Q^{d})
+z_{37}\,\alpha^{-,i}_{ab}(\partial Q^{a}Q^{b})+z_{38}\,\alpha^{-,i}_{ab}(Q^{a}G^{b})
+z_{39}\,\alpha^{-,i}_{ab}(UQ^{a}Q^{b})\ ,\nonumber
\ea
where the constants $z_{1}, \dots, z_{39}$ are listed in Appendix~\ref{app:1x1-linear}.
Upon redefining the currents of $A_\gamma$ and the component fields as in 
(\ref{eq:Ttilde}) -- (\ref{eq:Atilde}) and (\ref{eq:Vstilde}) -- (\ref{eq:V2}), respectively, 
these OPEs take the same form as in the non-linear calculation, i.e., as 
in (\ref{eq:ope2}) and (\ref{eq:ope3}), 
\ba
\widetilde V^{(1)}_{0}\,\widetilde V^{(1)}_{0} &\sim& \frac{1}{z^{2}}\,n_{1}\,\mathbb I\ , \\
\widetilde V^{(1)}_{0}\,\widetilde V^{(1), a}_{1/2} &\sim& \frac{1}{z}\,z_{1}\,\widetilde G^{a}\ ,\\
\widetilde V^{(1), a}_{1/2}\,\widetilde V^{(1), b}_{1/2} &\sim& \frac{1}{z^{3}}\,z_{6}\,\delta^{ab}\,\mb I
+\frac{1}{z^{2}}\,\Big(z_{7}\,\alpha^{-,i}_{ab}\widetilde A^{-,i}
+z_{8}\,\alpha^{+,i}_{ab}\widetilde A^{+,i}\Big)\nonumber \\
&&
+\frac{1}{z}\,\Big[
z_{11}\,\delta^{ab}(\widetilde A^{-,i}\,\widetilde A^{-,i})
+z_{12}\,\alpha^{-,i}_{ac}\alpha^{+,j}_{cb}\,(\widetilde A^{-,i}\,\widetilde A^{+,j})
+z_{13}\,\delta^{ab}(\widetilde A^{+,i}\,\widetilde A^{+,i})\nonumber \\
&& +z_{14}\,\alpha^{-,i}_{ab}\partial \widetilde A^{-,i}
+z_{21}\,\alpha^{+,i}_{ab}\partial \widetilde A^{+,i}+z_{26}\,\delta^{ab}\, \widetilde T
\Big]\ , \label{z8} \\
\widetilde V^{(1)}_{0}\,\widetilde V^{(1), +, i}_{1} &\sim& \frac{1}{z^{2}}\,z_{28}\,\widetilde A^{+,i}\ , \\
\widetilde V^{(1)}_{0}\,\widetilde V^{(1), +, i}_{1} &\sim& \frac{1}{z^{2}}\,z_{34}\,\widetilde A^{-,i} \ .
\ea
Indeed, comparing the relevant coefficients leads to 
\begingroup
\renewcommand*{\arraystretch}{1.3}
\be\label{tablecoeff}
\begin{array}{|cc|c|c|}
\hline
\mbox{OPE} & \mbox{field} & \widetilde A_{\gamma} & A_{\gamma} \\
\hline
V^{(1)}_0  \times V^{(1)}_{1/2} & \widetilde G^{a} & \frac{w_{1}}{n_{1}} 
= -\frac{\tilde k_{+}+\tilde k_{-}+2}{2\,\tilde k_{+}\,\tilde k_{-}} & \frac{z_{1}}{n_{1}} 
= -\frac{k_{+}+k_{-}}{2\,(k_{+}-1)\,(k_{-}-1)} \\
\hline
V^{(1)}_0  \times V^{(1)}_1 & \widetilde A^{+,i} & \frac{w_{2}}{n_{1}} 
= \frac{4}{\tilde k_{+}} & \frac{z_{28}}{n_{1}} = -\frac{4}{k_{+}-1} \\
\hline
V^{(1)}_{1/2}  \times V^{(1)}_{1/2} & \widetilde A^{+,i} & \frac{w_{15}}{n_{1}} 
= -\frac{4}{\tilde k_{+}} & 
\frac{z_{8}}{n_{1}} = \frac{4}{k_{+}-1} \\
\hline
V^{(1)}_{1/2}  \times V^{(1)}_{1/2} & (\widetilde A^{+,i}\,\widetilde A^{+,i}) & 
\frac{w_{11}}{n_{1}} = -\frac{1}{\tilde k_{+}\,\tilde k_{-}} & \frac{z_{11}}{n_{1}} 
= -\frac{1}{(k_{+}-1)(k_{-}-1)} \\
\hline
V^{(1)}_{1/2}  \times V^{(1)}_{1/2} & \widetilde T & 
\frac{w_{10}}{n_{1}} = \frac{2(\tilde k_{+}+\tilde k_{-}+2)}{\tilde k_{+}\,\tilde k_{-}} & 
\frac{z_{26}}{n_{1}} = -\frac{k_{+}+k_{-}}{(k_{+}-1)(k_{-}-1)} \\
\hline
\end{array}
\ee
\endgroup
Here $\tilde{k}_\pm$ are the levels of the non-linear realisation that are related to the levels $k_\pm$
of the linear realisation as in (\ref{eq:levelrel}). The ratios (that are normalisation independent)
match precisely once the various signs and factors of $2$ (that are a consequence of the different
conventions we have employed for the two calculations) have been taken into account.
For example, the normalisation
of the supercharges differs effectively by a factor of $\sqrt{2} i$, as follows by comparing 
eq.~(\ref{eq:GGans}) to eq.~(\ref{GGlin}).\footnote{In addition, there is a change of basis since we have used
a $\su(2)$-bispinor notation in the non-linear analysis, while for the linear analysis we have worked with
$\mathfrak{so}(4)$ vectors.} This also leads to a similar rescaling of the $V^{(s)}_{1/2}$ 
components of the multiplets.  Furthermore, for example the coefficient $z_8$ in (\ref{z8})
multiplies a matrix, which differs be a normalisation factor from the corresponding
matrix in (\ref{eq:ope3}) that is multiplied by $w_{15}$. Taking all of these factors carefully into account, 
the two calculations match exactly.
\medskip

Another example comes from the OPE of the first and second multiplet. Up to the level we considered
only the $V^{(1)}$ multiplet appears in the OPE, and for example, the Jacobi identities of the linear
${\cal W}_\infty$ algebra predict that we have 
\be
\widetilde{V}^{(1),a}_{1/2}\,\widetilde{V}^{(2)}_{0} \sim \frac{1}{z}\,\Phi^{a}_{5/2}\ ,
\ee
where $\Phi^{a}_{5/2}$ is an operator of conformal dimension $5/2$, transforming in the $(\frac{1}{2},\frac{1}{2})$ 
representation of $\su(2)\oplus\su(2)$. Its explicit form turns out to be 
\ba
\label{eq:Phi52}
\Phi^{a}_{5/2} &=&
w_{1}\,V^{(1),a}_{3/2}+w_{2}\,\partial V^{(1),a}_{1/2}+
w_{3}\,(\partial Q^{a}V^{(1)}_{0})+w_{4}\,(Q^{a}\partial V^{(1)}_{0})\nonumber \\
&&+w_{5}\,(G^{a}V^{(1)}_{0})+
w_{6}\,(UV^{(1),a}_{1/2})+w_{7}\,(U Q^{a}V^{(1)}_{0})\nonumber \\
&& +w_{8}\,\alpha^{+,i}_{ab}\,(A^{+,i}V^{(1),b}_{1/2})
+w_{9}\,\alpha^{-,i}_{ab}\,(A^{-,i}V^{(1),b}_{1/2})+\nonumber \\
&&
+w_{10}\,\alpha^{+,i}_{ab}\,(A^{+,i}Q^{b}V^{(1)}_{0})
+w_{11}\,\alpha^{-,i}_{ab}\,(A^{-,i}Q^{b}V^{(1)}_{0})\nonumber \\
&&+w_{12}\,\alpha^{+,i}_{ab}\,(Q^{b}V^{(1), +, i}_{1})
+w_{13}\,\alpha^{-,i}_{ab}\,(Q^{b}V^{(1),-,i}_{1})
+w_{14} \epsilon_{abcd}(Q^{b}Q^{c}Q^{d}V^{(1)}_{0})
\nonumber \\
&&
+w_{15}\,\alpha^{+,i}_{ab}\,\alpha^{+,i}_{cd}(Q^{c}Q^{d}V^{(1), b}_{1/2})
+w_{16}\,\alpha^{-,i}_{ab}\,\alpha^{-,i}_{cd}(Q^{c}Q^{d}V^{(1), b}_{1/2})\ ,
\ea
where the values of the coefficients are given explicitly in Appendix~\ref{app:1x2-linear}.
In terms of the decoupled fields we can write $\Phi^{a}_{5/2}$ as 
\ba
\Phi^{a}_{5/2} &=& w_{1}'\,\widetilde V^{(1),a}_{3/2}+w_{2}'\,\partial\widetilde V^{(1),a}_{1/2}+
w_{3}'\,(\widetilde G^{a}\,\widetilde V^{(1)}_{0})\nonumber \\
&& \qquad +w_{4}'\,\alpha^{+,i}_{ab}\,(\widetilde A^{+,i}\,\widetilde V^{(1),b}_{1/2})
+w_{5}'\,\alpha^{-,i}_{ab}\,(\widetilde A^{-,i}\,\widetilde V^{(1),b}_{1/2})\ ,
\ea
where
\be
w_{1}'=w_{1},\quad
w_{2}'=w_{2},\quad
w_{3}'=w_{5},\quad
w_{4}'=w_{8}+\frac{8\,w_{1}}{k_{+}+k_{-}},\quad
w_{5}'=w_{9}-\frac{8\,w_{1}}{k_{+}+k_{-}}\ .
\ee
In fact, this expression is (for generic coefficients $w'$) the most general solution of the decoupling conditions
\ba
U(z)\, \Phi_{5/2}^{a}(w) \sim 0\ , \qquad 
Q^{a}(z)\, \Phi_{5/2}^{a}(w) \sim 0\ .
\ea
We can finally bring it into the same form as the corresponding formula in (\ref{eq:ope7h})
using the $[\cdots]$ bracket
\ba
\Phi^{a}_{5/2} &=& w_{1}'\,\widetilde V^{(1),a}_{3/2}+
w_{3}'\,\left[\widetilde V^{(1)}_{0}\,\widetilde G^{a}\right]
 +w_{4}'\,\alpha^{+,i}_{ab}\,\left[\widetilde A^{+,i}\,\widetilde V^{(1),b}_{1/2}\right]
 +w_{5}'\,\alpha^{-,i}_{ab}\,\left[\widetilde A^{-,i}\,\widetilde V^{(1),b}_{1/2}\right]\ .\qquad 
\ea
As regards the structure constants, it only makes sense to compare ratios since the normalisation of 
$V^{(2)}_{0}$ is arbitrary. For example, the coefficient of 
$\left[\widetilde V^{(1)}_{0}\,\widetilde G^{a}\right]$ relative to $\widetilde V^{(1),a}_{3/2}$
equals (see Appendix~\ref{app:1x2-linear})
\ba
\frac{w_{3}'}{w_{1}'} &=& -\frac{16 (2 \gamma -1) (c (c+6)+18 (\gamma -1) \gamma )}{36 (c+2) \gamma ^2-36 (c+2) \gamma
   +c (24-(c-4) c)}\nonumber \\
   &=& \frac{4 \left(k_--k_+\right) \left(2 k_+ k_-+2 k_-+2 k_+-1\right)}{3 k_+^2 k_-^2-2 k_+
   k_-^2-2 k_-^2-2 k_+^2 k_--k_+ k_-+k_--2 k_+^2+k_+}\ .
\ea
This must be compared with the analogous quantity in the non-linear computation which is 
\ba
\frac{w_{50}}{w_{48}} &=&\frac{4 (\tilde k_--\tilde k_+) 
(2 \tilde k_+ \tilde k_-+4 \tilde k_-+4 \tilde k_++5)}{3 \tilde k_+^2 \tilde k_-^2
+4 \tilde k_+
 \tilde k_-^2-\tilde k_-^2+4 \tilde k_+^2 \tilde k_-+3 \tilde k_+ \tilde k_-
 -4 \tilde k_--\tilde k_+^2-4 \tilde k_+-4}\ ,
\ea
and one checks that they agree precisely, using again the relation between the levels (\ref{eq:levelrel}). 
Similarly, the coefficient of 
of $\left[\widetilde A^{+,i}\,\widetilde V^{(1),b}_{1/2}\right]$ relative to $\widetilde V^{(1),a}_{3/2}$ reads
\ba\label{w4p}
\frac{w_{4}'}{w_{1}'} &=& -\frac{64 \gamma  (c (\gamma -2)-6 (\gamma -1) \gamma ) ((c-6) c-18 (\gamma -1) 
(2 \gamma -1))}{c \left(-36 (c+2) \gamma ^2+36 (c+2) \gamma +c ((c-4) c-24)\right)}\nonumber \\
   &=& \frac{16 \left(k_-+2 k_+-1\right) \left(2 k_+ k_-^2-2 k_-^2-2 k_+
   k_-+k_--k_+\right)}{\left(k_-+k_+\right)\! \left(3 k_+^2 k_-^2-2 k_+ k_-^2-2 k_-^2-2 k_+^2
   k_--k_+ k_-+k_--2 k_+^2+k_+\right)}\ , \quad
\ea
and this matches precisely, using  eq.~(\ref{eq:levelrel}), the analogous ratio in the non-linear computation 
(again there is a relative factor of $-2$ because of the different conventions that were used, see the comment
below (\ref{tablecoeff}))
\be
-2\frac{w_{52}}{w_{48}} = \frac{16 (\tilde k_-+2 \tilde k_++2) 
(2 \tilde k_+ \tilde k_-^2+2 \tilde k_+
   \tilde k_--\tilde k_--\tilde k_+-2)}{(\tilde k_-+\tilde k_+
   +2) (3 \tilde k_+^2 \tilde k_-^2+4 \tilde k_+ \tilde k_-^2-\tilde k_-^2
   +4 \tilde k_+^2
   \tilde k_-+3 \tilde k_+ \tilde k_--4 \tilde k_--\tilde k_+^2-4 \tilde k_+-4)} \ .
\ee
The analysis for the coefficient of $\left[\widetilde A^{-,i}\,\widetilde V^{(1),b}_{1/2}\right]$ is the same 
since it can be obtained from (\ref{w4p}) upon exchanging $k_{\pm}\to k_{\mp}$.
\smallskip

These comparisons give rise to pretty non-trivial consistency checks of our analysis,
and it is very satisfying that they work out precisely. In summary, the results 
of this and the previous sections therefore give strong indications that
the ${\cal N}=4$ superconformal ${\cal W}_\infty$ algebra consisting of the large ${\cal N}=4$ superconformal
algebra as well as one ${\cal N}=4$ supermultiplet for each integer spin, are uniquely characterised in 
terms of the levels of the two $\su(2)$ algebras. This statement applies both to the linear as well as the 
non-linear version of the algebra.

\section{Conclusions}

In this paper we have studied the structure of the most general large ${\cal N}=4$
superconformal ${\cal W}_\infty$ algebra that contains, in addition to the superconformal
algebra, one ${\cal N}=4$ multiplet for each integer spin $s=1,2,\ldots$. We have found
strong evidence in favour of the claim that this family of algebras is uniquely characterised
in terms of the levels of the two $\su(2)$ algebras (that are a part of the 
large ${\cal N}=4$ superconformal algebra). Among other things, this shows that the
Wolf space cosets account essentially for all such ${\cal W}_\infty$ algebras. While 
this is natural from the perspective of these cosets, it is a little surprising that the 
complete structure of the algebra is essentially fixed by the large ${\cal N}=4$ algebra
itself --- this is to be compared with, say, the bosonic situation where the free parameter corresponding
to $\lambda$ encodes how the different (conformal) multiplets couple to one another.

Another consequence of this analysis is that the quantisation of the higher spin theory
is essentially unique. Indeed, both levels $k_\pm$ can be identified with parameters
of the (classical) higher spin theory, 
\be
 \lambda = \frac{k_-}{k_++k_-+2} \ , 
\quad \hbox{and} \quad
c = \frac{6k_+ k_- }{k_+ + k_- +2} \ ,
\ee
where $\lambda$ is the parameter that appears in the underlying higher spin algebra 
$\mathfrak{shs}_2[\lambda]$, while $c$ is the central charge that is determined in terms of the 
size of the AdS space. Note that our result is compatible with the explicit analysis of 
\cite{Gaberdiel:2014yla} where the asymptotic symmetry algebra of the higher spin 
theory was matched with the 't~Hooft limit of the Wolf space cosets --- both are the
't~Hooft  limit of a unique quantum ${\cal W}_{\infty}$ algebra, and hence must agree.

In the limit where one of the levels of the two $\su(2)$ algebras goes to infinity, the large
${\cal N}=4$ superconformal algebra can be truncated to the small ${\cal N}=4$ 
superconformal algebra. Thus our analysis predicts that there is at least one family
of ${\cal W}_{\infty}$ algebras with small ${\cal N}=4$ superconformal algebra
that are labelled by the level of the surviving $\su(2)$ algebra (or equivalently by the
central charge). It would be interesting to see whether this accounts for all
small ${\cal N}=4$ algebras with this multiplet spectrum, or whether there are additional
constructions that cannot be obtained as a limit of a large ${\cal N}=4$ superconformal
${\cal W}$ algebra. In particular, one may expect that the ${\cal W}$ algebra that is relevant 
for string theory on ${\rm AdS}_3 \times {\rm S}^3 \times {\rm K3}$ should not appear in
this fashion.

\section*{Acknowledgements}

We thank Cheng Peng and Carl Vollenweider for useful discussions. The research
of CC and MRG is supported in parts by the Swiss National Science Foundation.

\appendix

\section{The Structure of the Supermultiplet}\label{app:supmult}

In this appendix we specify our conventions for the OPEs of the superconformal generators with the 
various component fields of the ${\cal N}=4$ supermultiplet. 

\subsection{The Non-linear Case}

For the case of the non-linear $\tilde{A}_\gamma$ algebra, the OPEs of the stress-energy tensor and 
the affine currents were given already in eqs.~(\ref{Tans}) and (\ref{Aans}). Our ansatz for the OPEs of the 
supercharges with the component fields of the ${\cal N}=4$ supermultiplet is 
\begin{align*}
G^{\alpha\beta}(z)V^{(s)}_0(w) &\sim \frac{V^{(s)\alpha\beta}_{1/2}(w)}{z-w}\ ,\\ 
G^{\alpha\beta}(z)V^{(s)\gamma\delta}_{1/2}(w) &\sim\frac{ g_{1/2,1}\epsilon_{\alpha\gamma}\epsilon_{\beta\delta}V^{(s)}_0(w)}{(z-w)^2}+ \frac{1}{z-w}\left[g_{1/2,2} \epsilon_{\alpha\gamma}\epsilon_{\beta\delta}\partial V^{(s)}_0(w)+ {} 
\right.\\
{}& \left. {}+ \epsilon_{\beta\delta}\ell_{i,\alpha\gamma} V^{(s)+i}_{1}(w)+ \epsilon_{\alpha\gamma}\ell_{i,\beta\delta}V^{(s)-i}_{1}(w)\right]\ ,\\ 
G^{\alpha\beta}(z)V^{(s)+ i}_1(w)&\sim \frac{g^+_{1,1} \rho^i_{\gamma\alpha} V^{(s)\gamma\beta}_{1/2}(w)}{(z-w)^2}+ \frac{1}{z-w}\left\{
 \rho^i_{\gamma\alpha} \big[g^+_{1,2} \partial V^{(s)\gamma\beta}_{1/2}(w) +{}\right.\\
&\left. {}+ 
g^+_{1,3}\ell_{j,\delta\beta} (A^{-j}V^{(s)\gamma\delta}_{1/2})(w) +  V^{(s)\gamma\beta}_{3/2}(w) \big] + g^+_{1,4}(A^{+i}V^{(s)\alpha\beta}_{1/2})(w) \right\}
\ ,\\
G^{\alpha\beta}(z)V^{(s)- i}_1(w)&\sim \frac{g^-_{1,1} \rho^i_{\gamma\beta} V^{(s)\alpha\gamma}_{1/2}(w)}{(z-w)^2}+ \frac{1}{z-w}\left\{
 \rho^i_{\gamma\beta} \big[g^-_{1,2} \partial V^{(s)\alpha\gamma}_{1/2}(w) +{}\right.\\
&\left. {}+ g^-_{1,3}\ell_{j,\delta\alpha} (A^{+j}V^{(s)\delta\gamma}_{1/2})(w) +g_{1,5} V^{(s)\alpha\gamma}_{3/2}(w) \big] + g^-_{1,4}(A^{-i}V^{(s)\alpha\beta}_{1/2})(w) \right\}
\ ,\\
G^{\alpha\beta}(z)V^{(s)\gamma\delta}_{3/2}(w)&\sim \frac{g_{3/2,1}\epsilon_{\alpha\gamma}\epsilon_{\beta\delta}V^{(s)}_0(w)}{(z-w)^3}+
\frac{1}{(z-w)^2}\left[\epsilon_{\beta\delta}\ell_{i,\alpha\gamma}g^+_{3/2,2}V^{(s)+ i}_1(w)+ {}\right.\\
&\left.{}+\epsilon_{\alpha\gamma}\ell_{i,\beta\delta}g^-_{3/2,2}V^{(s)- i}_1(w)\right]+
\frac{1}{z-w}\left\{
\epsilon_{\beta\delta}\ell_{i,\alpha\gamma}\big[g^+_{3/2,3}(\partial A^{+i}V^{(s)}_0)(w)+ {}\right.\\
&\left.{}+ g^+_{3/2,4} (A^{+i}\partial V^{(s)}_0)(w)+ g^+_{3/2,5}\partial V^{(s)+ i}_1(w)\big] + {}\right.\\
{}&{}+ \epsilon_{\alpha\gamma}\ell_{i,\beta\delta}\big[g^-_{3/2,3}(\partial A^{-i}V^{(s)}_0)(w) +g^-_{3/2,4} (A^{-i}\partial V^{(s)}_0)(w) + {}\\
{}&{}+ g^-_{3/2,5}\partial V^{(s)- i}_1(w)\big]+ {f^i}_{jl} \big[g^+_{3/2,6}\epsilon_{\beta\delta}\ell_{i,\alpha\gamma}(A^{+j}V^{(s)+l}_1)(w)+{}\\
&{}+g^-_{3/2,6}\epsilon_{\alpha\gamma}\ell_{i,\beta\delta}(A^{-j}V^{(s)-l}_1)(w)\big]+ \epsilon_{\alpha\beta}\epsilon_{\gamma\delta}V^{(s)}_2(w)+{}\\
{}&\left. {}+\big( g^+_{3/2,7}\epsilon_{\beta\delta}\epsilon_{\nu\sigma}\ell_{i,\alpha\gamma}r^i_{\mu\rho}+ g^-_{3/2,7}\epsilon_{\alpha\gamma}\epsilon_{\mu\rho}\ell_{i,\beta\delta}r^i_{\nu\sigma}\big)(G^{\mu\nu}V^{(s)\rho\sigma}_{1/2})(w) \right\}\ ,
\\
G^{\alpha\beta}(z)V^{(s)}_{2}(w)&\sim \frac{g_{2,1}V^{(s)\alpha\beta}_{1/2}(w)}{(z-w)^3}+
\frac{1}{(z-w)^2}\left[g_{2,2}V^{(s)\alpha\beta}_{3/2}(w)+ g^+_{2,3}\rho_{i,\gamma\alpha}(A^{+i}V^{(s)\gamma\beta}_{1/2})(w)+{}\right.\\
&\left. {}+g^-_{2,3}\rho_{i,\gamma\beta}(A^{-i}V^{(s)\alpha\gamma}_{1/2})(w)\right]+ \frac{1}{z-w}\left[
g_{2,4}\partial V^{(s)\alpha\beta}_{3/2}(w) + {}\right.\\
&\left.{}+\rho_{i,\gamma\alpha} g^+_{2,5}(\partial A^{+i} V^{(s)\gamma\beta}_{1/2})(w)+ g^-_{2,5}(\partial A^{-i} V^{(s)\alpha\gamma}_{1/2})(w)\right]\ .
\end{align*}
Here $(AB)$ denotes the minimal normal ordering of 2 operators $A$ and $B$, i.e.\ the regular term in the OPE 
between $A$ and $B$, and the matrices $r^i$ are defined via
\begin{equation}
r^i_{\alpha\beta} = \rho^i_{\alpha\gamma}\epsilon_{\gamma\beta}\ ,
\end{equation}
i.e., $r^i$ it is the matrix $\rho^i$ with one of the $(\alpha,\beta)$ indices raised.
With this ansatz, the Jacobi identities with the $\mathcal{N}=4$ superconformal algebra fix the undetermined structure constants uniquely, and we find in addition to eqs.~(\ref{c1}) and (\ref{c2}) the values
\begin{align}\label{c3}
&g_{1/2,1} = -4s\ ,\quad  g_{1/2,2} = -2\ , \quad
g^\pm_{1,1} = -\frac{4 [-1 + k_\pm  + s(2   + k_+ +  k_-)]}{2 + k_+ + k_-}\ , \notag \\ 
&g^\pm_{1,2}=-\frac{4 [1 + k_\pm +s(2  + k_+  + k_-)]}{(1 + 2 s) (2 + k_+ + k_-)}\ ,\quad
g^\pm_{1,3}=-2g^\pm_{1,4}=\frac{8}{2 + k_+ + k_-}\ ,\quad
g_{1,5}= -1\ , \notag \\
&g_{3/2,1}=-\frac{32 s (1 + s) (k_+ - k_-)}{(1 + 2 s) (2 + k_+ + k_-)}\ ,\quad g^\pm_{3/2,2} = \mp \frac{8 (1 + s) (1 + k_\mp +s(2  + k_+  + k_-)]}{(1 + 2 s) (2 +     k_+ + k_-)}\ , \notag\\
&g^\pm_{3/2,3}= \frac{\pm 16 s}{2 + k_+ + k_-}\ ,\quad 
g^\pm_{3/2,4}= \frac{\mp16}{2 + k_+ + k_-}\ ,\quad g^\pm_{3/2,5}=  \mp\frac{4 [1 +k_\mp +s(2  + k_+ + k_- )]}{(1 + 2 s) (2 + k_+ + k_-)}\ ,
\notag \\
&g^\pm_{3/2,6} =-g^\pm_{3/2,7} = \frac{\mp 4}{2 + k_+ + k_-}\ ,\quad 
g_{2,1} = \frac{32 s (1 + s) (k_+ - k_-)}{(1 + 2 s) (2 + k_+ + k_-)}\ ,\quad g_{2,2} = -2 (3 + 2 s)\ , \notag \\
& g^\pm_{2,3}=\pm \frac{ 32 (1 + s)}{2 + k_+ + k_-}\ ,\quad 
g_{2,4}=-2\ ,\quad 
g^\pm_{2,5}= \pm \frac{16 (1 + s)}{2 + k_+ + k_-}\ . 
\end{align}
\bigskip

\subsection{The Linear Case}\label{app:linsup}

In this subsection we explain our conventions for the OPEs of the fields of the linear
$A_\gamma$ algebra with the component fields of the supermultiplet. For the component
fields we use the conventions explained in eq.~(\ref{lincomp}).
\begin{align}
A_{\gamma}\times V_{0}^{(s)}: && U\,V_{0}^{(s)} &\sim 0\ ,\quad Q^{a}\,V_{0}^{(s)} \sim 0\ ,\quad A^{\pm,i}\,V_{0}^{(s)} \sim0\ , \quad G^{a}\,V_{0}^{(s)} \sim \frac{1}{z}\,V_{1/2}^{(s), a}\ , \notag
\end{align}
%
\begin{align}\notag
A_{\gamma}\times V_{1/2}^{(s), a}:&& U\,V_{1/2}^{(s), a} &\sim 0\ ,\quad
Q^{a}\,V_{1/2}^{(s), b} \sim 0\ , \quad
A^{\pm,i}\,V_{1/2}^{(s), a} \sim \frac{1}{z}\,\alpha_{ab}^{\pm,i}\,V_{1/2}^{(s), b}\ ,\\
&&G^{a}\,V_{1/2}^{(s), b} &\sim \frac{2s\,\delta^{ab}}{z^{2}}\,V_{0}^{(s)}+\frac{1}{z}\left(
\alpha_{ab}^{+,i}\,V_{1}^{(s), +,i}+\alpha_{ab}^{-,i}\,V_{1}^{(s), -,i}+\delta^{ab}\,
\partial V_{0}^{(s)} \right)\ , \notag 
\end{align}
%
\begin{align}
A_{\gamma}\times V_{1}^{(s), \pm, i}:&& U\,V_{1}^{(s),\pm, i} &\sim 0\ , \quad Q^{a}\,V_{1}^{(s), \pm,i} \sim \pm\frac{2}{z}\,\alpha^{\pm, i}_{ab}\,V_{1/2}^{(s), a}\ , \notag \\
&& A^{\pm ,i}\,V_{1}^{(s),\pm, j} &\sim \frac{2s}{z^{2}}\,\delta^{ij}\,V_{0}^{(s)}
+\frac{1}{z}\,\epsilon^{ijk}\,V_{1}^{(s), \pm, k}\ ,  \quad A^{\pm,i}\,V_{1}^{(s), \mp,j} \sim0\ , \notag \\
&& G^{a}\,V_{1}^{(s), \pm, i} &\sim 4(s+\gamma_\mp)\Big(\frac{1}{z^{2}}+\frac{1}{z(2s+1)}\, \partial\Big)\,\alpha^{\pm, i}_{ab}\,V_{1/2}^{(s),b}
\mp \frac{1}{z}\,\alpha^{\pm, i}_{ab} V_{3/2}^{(s),b}\ ,  \notag 
\end{align}
%
\begin{align}
A_{\gamma}\times V_{3/2}^{(s), a}:&& U\,V_{3/2}^{(s), a} &\sim -\frac{1}{z^{2}}\,V_{1/2}^{(s), a}\ , \notag\\
&&Q^{a}\,V_{3/2}^{(s), b} &\sim \frac{4s \delta^{ab}}{z^{2}}\,V_{0}^{(s)}  +\frac{2}{z}\,\left(\alpha_{ab}^{+,i}\,V_{1}^{(s), +,i}+ \alpha_{ab}^{-,i}\,V_{1}^{(s), -,i} 
- \delta^{ab}\,\partial V_{0}^{(s)}
\right)\ ,\nonumber \\
&& A^{\pm ,i}\,V_{3/2}^{(s), a} &\sim \pm \frac{8s(s+1)+\gamma_\mp}{z^2(2s+1)} \alpha^{\pm,i}_{ab}\,V_{1/2}^{(s), b}+\frac{1}{z}\,\alpha_{ab}^{\pm,i}\,V_{3/2}^{(s), b}\ ,\notag \\
&& G^{a}\,V_{3/2}^{(s), b} &\sim
-\frac{16s(s+1)(2\gamma-1)}{z^3(2s+1)} \delta^{ab}\,V_{0}^{(s)} - \frac{8(s+1)}{(2s+1)}\Big(\frac{1}{z^{2}}+\frac{1}{2(s+1)z}\,\partial \Big)\times {} \notag\\
 && &\quad {}\times \Big[(s+ \gamma_+)\alpha^{+,i}_{ab}\,V_{1}^{(s), +,i}
-(s+\gamma_-) \alpha^{-,i}_{ab}\,V_{1}^{(s), -,i}
\Big] +\frac{1}{z}\, \delta^{ab}\,V_{2}^{(s)}
\notag \ .
\end{align}
\begin{align}
A_{\gamma}\times V_{2}^{(s)}:&& U\,V_{2}^{(s)} &\sim \frac{8\,s}{z^{3}}\,V_{0}^{(s)}-\frac{4}{z^{2}}\,\partial V_{0}^{(s)}\ , \notag \\
&&Q^{a}\,V_{2}^{(s)} &\sim -\frac{2(2s+1)}{z^{2}}\,V_{1/2}^{(s), a}+\frac{2}{z}\,\partial V_{1/2}^{(s), a}\ , \notag \\ 
&&A^{\pm,i}\,V_{2}^{(s)} &\sim \frac{\pm2(s+1)}{z^{2}}\,V_{1}^{(s), \pm,i}\ , \notag \\
&&G^{a}\,V_{2}^{(s)} &\sim \frac{16(2\gamma-1)\,s(s+1)}{z^{3}(2s+1)} \,V_{1/2}^{(s), a}+\frac{2s+3}{z^{2}}\,V_{3/2}^{(s), a}+\frac{1}{z}\,\partial V_{3/2}^{(s), a}\ , \notag \\
&&T\,V_{2}^{(s)} &\sim -\frac{24(2\gamma-1)s(s+1)}{z^{4}(2s+1)}\,V_{0}^{(s)}+\frac{s+2}{z^{2}}\,V_{2}^{(s)}+\frac{1}{z}\,\partial V_{2}^{(s)}\ . \notag
\end{align}
Here $\gamma_+=\gamma$ and $\gamma_-=\bar\gamma=1-\gamma $.
Only the component field $V_{2}^{(s)}$ is quasi-primary (but not primary).

\section{The Spin 4 OPEs and the Structure Constants} \label{sec:A}

\subsection{The OPEs}
\noindent The general ansatz for the OPEs of total spin $4$ is
\begin{align} \notag
V^{(1)}_{0}\times V^{(3)}_0 \sim{}&
w_{80} T + w_{81} V^{(2)}_0  + w_{82} [V^{(1)}_0V^{(1)}_0] + w_{83} [A^{+i}{A^{+}}_i] + w_{84} [A^{-i}{A^{-}}_i] + {}\\\notag&{}+
w_{85} V^{(3)}_0 +  w_{86} [V^{(1)}_0 V^{(2)}_0] + w_{87} V^{(1)}_2 + w_{88} [V^{(1)}_0 [V^{(1)}_0V^{(1)}_0]]+ {}\\\notag&{}+ w_{89} [T V^{(1)}_0]+ 
\epsilon_{\alpha\gamma}\epsilon_{\beta\delta} w_{90} [G^{\alpha\beta}V^{(1)\gamma\delta}_{1/2}]+
w_{91} [{A^+}_i V^{(1)+i}_1]+ {}\\\notag &{}+ w_{92} [{A^-}_i V^{(1)-i}_1] + w_{93} [[{A^+}_i A^{+i}]V^{(1)}_0] +
w_{94} [[{A^-}_i A^{-i}]V^{(1)}_0] \ ,\\\notag
V^{(1)}_{0}\times V^{(1)}_2 \sim{}&
w_{95} T +
\cdots +
w_{108} [[{A^+}_i A^{+i}]V^{(1)}_0] + w_{109} [[{A^-}_i A^{-i}]V^{(1)}_0] \ ,\\\notag
V^{(1)}_0 \times V^{(2)+i}_1 \sim {}&
w_{110} A^{+i}+
w_{111} V^{(1)+ i}_1 + w_{112} [A^{+ i} V^{(1)}_0] +
w_{113} V^{(2)+i}_1 + {}\\\notag&{}+ w_{114} [V^{(1)}_0 V^{(1)+i}_1]+  w_{115} [A^{+ i} V^{(2)}_0]+ w_{116} [A^{+i}V^{(1)}_0]_{-1}+{}\\\notag&{}+
w_{117} [A^{+i}[V^{(1)}_0V^{(1)}_0]]+
w_{118} [TA^{+i}]+ \epsilon_{\beta\delta} r^i_{\alpha\gamma} \big( w_{119} [V^{(1)\alpha\beta}_{1/2}V^{(1)\gamma\delta}_{1/2}]+ {}\\\notag&{}+ w_{120} [G^{\alpha\beta}V^{(1)\gamma\delta}_{1/2}]+
w_{121} [G^{\alpha\beta}G^{\gamma\delta}] \big)+ 
{f^{i}}_{jl} \big( w_{122} [A^{+j}V^{(1)+l}_1]+ {}\\\notag&{}+ w_{123} [A^{+j}A^{+l}]_{-1} \big) +  w_{124} [[{A^+}_jA^{+j}]A^{+i}]+w_{125} [[{A^-}_jA^{-j}]A^{+i}]\ ,\\\notag
V^{(1)+i}_1 \times V^{(2)}_0 \sim {}& 
w_{126}A^{+i}+
\cdots +
w_{140} [[{A^+}_jA^{+j}]A^{+i}]+ w_{141} [[{A^-}_jA^{-j}]A^{+i}]\ ,\\\notag
V^{(1)}_0 \times V^{(2)-i}_1 \sim {}&
 w_{142}A^{-i}+
 w_{143}V^{(1)- i}_1 +  w_{144}[A^{- i} V^{(1)}_0] +
 w_{145} V^{(2)-i}_1 +  {}\\\notag&{}+
w_{146} [V^{(1)}_0 V^{(1)-i}_1]+  w_{147} [A^{- i} V^{(2)}_0]+w_{148}  [A^{-i}V^{(1)}_0]_{-1}+ {}\\\notag&{}+
w_{149} [A^{-i}[V^{(1)}_0V^{(1)}_0]]+w_{150} [TA^{-i}]+
\epsilon_{\alpha\gamma} r^i_{\beta\delta} \big( w_{151} [V^{(1)\alpha\beta}_{1/2}V^{(1)\gamma\delta}_{1/2}]+ {}\\\notag&{}+
w_{152} [G^{\alpha\beta}V^{(1)\gamma\delta}_{1/2}]+w_{153} [G^{\alpha\beta}G^{\gamma\delta}] \big)+ 
{f^{i}}_{jl} \big( w_{154} [A^{-j}V^{(1)-l}_1]+ {}\\\notag&{}+w_{155} [A^{-j}A^{-l}]_{-1} \big) + w_{156} [[{A^-}_jA^{-j}]A^{-i}]+w_{157} [[{A^+}_jA^{+j}]A^{-i}]\ ,\\\notag
V^{(1)-i}_1 \times V^{(2)}_0 \sim {}& 
w_{158} A^{-i}+
\cdots +
w_{172}[[{A^-}_jA^{-j}]A^{-i}]+ w_{173}[[{A^+}_jA^{+j}]A^{-i}]\ ,\\\notag
V^{(1)+ i }_1\times V^{(1)+ j }_1\sim {}&
\eta^{ij} \big( w_{174} I+ w_{175} T +  w_{176} V^{(2)}_0 +  w_{177} [V^{(1)}_0V^{(1)}_0 ]+  w_{178} [A^{+l}{A^{+}}_l] +  {}\\\notag&{}+ w_{179} [A^{-l}{A^{-}}_l] \big) + w_{180} [A^{+ i}A^{+ j}]+
{f^{ij}}_l\big\{ w_{181}A^{+ l} + w_{182}V^{(2)+ l}_1+{}\\\notag&{}+w_{183}[V^{(1)}_0 V^{(1)+ l}_1]+ w_{184} [A^{+ l}V^{(2)}_0]+    w_{185} [A^{+ l}V^{(1)}_0]_{-1}+{}\\\notag&{}+
 w_{186} [A^{+ l}[V^{(1)}_0V^{(1)}_0]] +  w_{187} [TA^{+ l}] + 
\epsilon_{\beta\delta}r^l_{\alpha\gamma} \big(  w_{188} [V^{(1)\alpha\beta}_{1/2}V^{(1)\gamma\delta}_{1/2}]+{}\\\notag&{}+ w_{189} [G^{\alpha\beta}V^{(1)\gamma\delta}_{1/2}]+  w_{190} [G^{\alpha\beta}G^{\gamma\delta}] \big)+ 
{f^l}_{pq} \big(  w_{191} [A^{+p}V^{(1)+q}_1]+ {}\\\notag&{}+w_{192} [A^{+p}A^{+q}]_{-1} \big) +   w_{193} [[{A^+}_pA^{+p}]A^{+l}]+ w_{194} [[{A^-}_pA^{-p}]A^{-l}]
\big\}\ ,\\\notag
V^{(1)- i }_1\times V^{(1)- j }_1\sim {}&
\eta^{ij} \big( w_{195} I +  w_{196} T +  w_{197} V^{(2)}_0 +  w_{198} [V^{(1)}_0V^{(1)}_0 ]+  w_{199} [A^{+l}{A^{+}}_l] + {}\\\notag&{}+ w_{200} [A^{-l}{A^{-}}_l] \big) + w_{201} [A^{- i}A^{- j}]+
{f^{ij}}_l\big\{  w_{202} A^{- l} + w_{203} V^{(2)- l}_1+ {}\\\notag&{}+w_{204} [V^{(1)}_0 V^{(1)- l}_1]+  w_{205} [A^{-l}V^{(2)}_0]+ w_{206} [A^{-l}V^{(1)}_0]_{-1}+{}\\\notag&{}+
 w_{207} [A^{- l}[V^{(1)}_0V^{(1)}_0]] +  w_{208} [TA^{- l}] + 
\epsilon_{\alpha\gamma}r^l_{\beta\delta} \big(  w_{209} [V^{(1)\alpha\beta}_{1/2}V^{(1)\gamma\delta}_{1/2}]+ {}\\\notag&{}+w_{210} [G^{\alpha\beta}V^{(1)\gamma\delta}_{1/2}]+  w_{211} [G^{\alpha\beta}G^{\gamma\delta}] \big)+
{f^l}_{pq} \big(  w_{212} [A^{-p}V^{(1)-q}_1]+{}\\\notag&{}+ w_{213} [A^{-p}A^{-q}]_{-1} \big) +   w_{214} [[{A^-}_pA^{-p}]A^{-l}]+ w_{215} [[{A^+}_pA^{+p}]A^{-l}]
\big\}\ ,\\\notag
V^{(1)+ i }_1\times V^{(1)- j }_1\sim {}&
w_{216}[A^{+i}A^{-j}] +r^i_{\alpha\gamma}r^j_{\beta\delta} w_{217}[G^{\alpha\beta}V^{(1)\gamma\delta}_{1/2}]+ w_{218}[A^{+i}V^{(1)-j}]+{}\\\notag&{}+w_{219}[A^{-j}V^{(1)+i}]+w_{220} [A^{+i}[A^{-j}V^{(1)}_0]]+w_{221}[A^{+i}A^{-j}]_{-1}\ ,\\\notag
V^{(1)\alpha\beta}_{1/2}\times V^{(1)\gamma\delta}_{3/2}\sim {}&
\epsilon_{\alpha\gamma}\epsilon_{\beta\delta} \big\{ w_{222} V^{(1)}_0 +  w_{223} T+  w_{224} V^{(2)}_0 +  w_{225} [V^{(1)}_0V^{(1)}_0]+  {}\\\notag &{}+ w_{226} [A^{+i}{A^{+}}_i] + w_{227} [A^{-i}{A^{-}}_i] + 
 w_{228} V^{(3)}_0 +  w_{229} [V^{(1)}_0 V^{(2)}_0]+  {}\\\notag &{}+ w_{230} V^{(1)}_2 +   w_{231}  [V^{(1)}_0[V^{(1)}_0V^{(1)}_0]]+   w_{232} [TV^{(1)}_0]+{}\\\notag &{}+
\epsilon_{\rho\mu}\epsilon_{\sigma\nu}  w_{233} [G^{\rho\sigma}V^{(1)\mu\nu}_{1/2}]+
  w_{234} [{A^+}_iV^{(1)+i}]+  w_{235} [{A^-}_iV^{(1)-i}]+ {}\\\notag &{}+
  w_{236}  [[{A^+}_iA^{+i}]V^{(1)}_0]+  w_{237}  [[{A^-}_iA^{-i}]V^{(1)}_0] \big\} +
\epsilon_{\beta\delta}\ell_{i,\alpha\gamma} \big\{  w_{238} A^{+i}+{}\\\notag &{}+  w_{239} V^{(1)+i}_1+  w_{240} [A^{+i}V^{(1)}_0] +
 w_{241} V^{(2)+i}_1 +  w_{242} [V^{(1)}_0V^{(1)+i}_1]+  {}\\\notag &{}+
  w_{243} [A^{+i}V^{(2)}_0]+  w_{244} [A^{+i}V^{(1)}_0]_{-1}+
 w_{245} [A^{+i}[V^{(1)}_0V^{(1)}_0]]+ {}\\\notag&{}+ w_{246} [TA^{+i}]+
\epsilon_{\sigma\nu}r^i_{\rho\mu} \big(  w_{247} [V^{(1)\rho\sigma}_{1/2}V^{(1)\mu\nu}_{1/2}]+ w_{248} [G^{\rho\sigma}V^{(1)\mu\nu}_{1/2}]+ {}\\\notag&{}+ w_{249} [G^{\rho\sigma}G^{\mu\nu}] \big)+ 
 {f^i}_{jl} \big(  w_{250} [A^{+j}V^{(1)+l}_1]+ w_{251} [A^{+j}A^{+l}]_{-1} \big) + {}\\\notag &{}+
 w_{252} [[{A^+}_jA^{+j}]A^{+i}]+  w_{253} [[{A^-}_jA^{-j}]A^{+i}] \big\}+ 
\epsilon_{\alpha\gamma}\ell_{i,\beta\delta} \big\{  w_{254} A^{-i}+{}\\\notag&{}+ w_{255} V^{(1)- i}_1 +  w_{256} [A^{-i}V^{(1)}_0]+  w_{257} V^{(2)-i}_1 +  w_{258} [V^{(1)}_0V^{(1)-i}_1]+  {}\\\notag&{}+
 w_{259} [A^{-i}V^{(2)}_0]+   w_{260} [A^{-i}V^{(1)}_0]_{-1}+
  w_{261} [A^{-i}[V^{(1)}_0V^{(1)}_0]]+ {}\\\notag &{}+ w_{262}  [TA^{-i}]+
\epsilon_{\rho\mu}r^i_{\sigma\nu} \big(   w_{263} [V^{(1)\rho\sigma}_{1/2}V^{(1)\mu\nu}_{1/2}]+  w_{264} [G^{\rho\sigma}V^{(1)\mu\nu}_{1/2}]+ {}\\\notag &{}+ w_{265} [G^{\rho\sigma}G^{\mu\nu}] \big) + 
{f^i}_{jl} \big(  w_{266}  [A^{-j}V^{(1)-l}_1]+  w_{267} [A^{-j}A^{-l}]_{-1} \big) +  {}\\\notag &{}+  w_{268} [[{A^-}_jA^{-j}]A^{-i}]+     w_{269}  [[{A^+}_jA^{+j}]A^{-i}]\big\}+ {}\\\notag&{}+
\ell_{i,\alpha\gamma}\ell_{i,\beta\delta} \big\{ w_{270} [A^{+i}A^{-j}]+
+ r^i_{\rho\mu}r^j_{\sigma\nu} w_{271} [G^{\rho\sigma}V^{(1)\mu\nu}_{1/2}]+ {}\\\notag &{}+  w_{272} [A^{+i}V^{(1)-j}_1]+  w_{273}  [A^{-j}V^{(1)+i}_1]+ w_{274} [A^{+i}[A^{-j}V^{(1)}_0]]+ {}\\\notag &{}+
 w_{275} [A^{+i}A^{-j}]_{-1}\big\}\ ,\\\notag
%
%
%
V^{(1)\alpha\beta}_{1/2}\times V^{(2)\gamma\delta}_{1/2}\sim {}&
\epsilon_{\alpha\gamma}\epsilon_{\beta\delta} \big\{  w_{276}  V^{(1)}_0 + w_{277} T+
\cdots +
w_{291}  [[{A^-}_iA^{-i}]V^{(1)}_0] \big\} +
\cdots+{}\\\notag&{}+
%
%
\ell_{i,\alpha\gamma}\ell_{i,\beta\delta} \big\{ w_{324} [A^{+i}A^{-j}]+
\cdots +
 w_{329} [A^{+i}A^{-j}]_{-1}\big\}\ ,\\\notag
V^{(2)}_0\times V^{(2)}_0 \sim {}& 
n_2 I + w_{330} T + w_{331} V^{(2)}_0 + w_{332}[V^{(1)}_0V^{(1)}_0]+ w_{333} [{A^+}_iA^{+i}]+ {}\\ &{}+w_{334} [{A^-}_iA^{-i}]\ .
\label{eq:ope4}
\end{align}
As in the main part of the paper, we have labelled the structure constants in the OPE 
$V^{(1)}_0\times V^{(1)}_2$ in the same order as in the OPE  $V^{(1)}_0\times V^{(3)}_0$ given 
above it, which is of the same form;
the structure constants in the OPE $V^{(1)+i}_1\times V^{(2)}_0$ in the same order as in the 
OPE $V^{(1)}_0\times V^{(2)+i}_1$ given above it, etc.

\subsection{The Structure Constants}\label{app:B1}


In this section we list the structure constants in the OPEs~\eqref{eq:ope4}. We have 
fixed the redefinition freedom of $V^{(2)}_0$ and $V^{(3)}_0$ with the 
conditions~(\ref{eq:red1}), (\ref{eq:red2}), and we have assumed the 
parity symmetry~\eqref{eq:parity} so that eq.~\eqref{eq:conj} holds. Under these assumptions the 
structure constants take  the following form:
\begin{align*}
 w_{95}&=-\tfrac{64 \left(k_--k_+\right)\left(-2-k_--k_++k_- k_+\right)\left(2+k_-+k_++k_- k_+\right)}{\left(2+k_-+k_+\right) \left(-4-4 k_--k_-^2-4
k_++3 k_- k_++4 k_-^2 k_+-k_+^2+4 k_- k_+^2+3 k_-^2 k_+^2\right)} \ ,\\
 w_{96}&=8 \ ,\\
 w_{97}&=\tfrac{16 \left(k_--k_+\right) \left(5+4 k_-+4 k_++2 k_- k_+\right)}{-4-4 k_--k_-^2-4 k_++3 k_- k_++4 k_-^2 k_+-k_+^2+4 k_- k_+^2+3 k_-^2 k_+^2}
\ ,\\
 w_{98}&=-\tfrac{32 \left(-1+k_-\right) \left(1+k_-\right) k_+ \left(2+2 k_-+k_+\right)}{\left(2+k_-+k_+\right) \left(-4-4 k_--k_-^2-4 k_++3 k_- k_++4
k_-^2 k_+-k_+^2+4 k_- k_+^2+3 k_-^2 k_+^2\right)} \ ,\\
 w_{99}&=\tfrac{32 k_- \left(-1+k_+\right) \left(1+k_+\right) \left(2+k_-+2 k_+\right)}{\left(2+k_-+k_+\right) \left(-4-4 k_--k_-^2-4 k_++3 k_- k_++4
k_-^2 k_+-k_+^2+4 k_- k_+^2+3 k_-^2 k_+^2\right)} \ ,\\
 w_{111}&=-\tfrac{16 \left(-1+k_-\right) \left(1+k_-\right) k_+ \left(1+k_+\right) \left(2+2 k_-+k_+\right)}{\left(2+k_-+k_+\right) \left(-4-4 k_--k_-^2-4
k_++3 k_- k_++4 k_-^2 k_+-k_+^2+4 k_- k_+^2+3 k_-^2 k_+^2\right)} \ ,\\
 w_{112}&=\tfrac{64 \left(-1+k_-\right) \left(1+k_-\right) \left(1+k_+\right) \left(2+2 k_-+k_+\right)}{\left(2+k_-+k_+\right) \left(-4-4 k_--k_-^2-4
k_++3 k_- k_++4 k_-^2 k_+-k_+^2+4 k_- k_+^2+3 k_-^2 k_+^2\right)} \ ,\\
 w_{120}&=\tfrac{8 \left(-1+k_-\right) \left(1+k_+\right) \left(2+2 k_-+k_+\right) \left(2+k_-+2 k_+\right)}{\left(2+k_-+k_+\right) \left(-4-4 k_--k_-^2-4
k_++3 k_- k_++4 k_-^2 k_+-k_+^2+4 k_- k_+^2+3 k_-^2 k_+^2\right)} \ ,\\
 w_{127}&=-\tfrac{16 \left(-1+k_-\right) \left(1+k_-\right) k_+ \left(1+k_+\right) \left(2+2 k_-+k_+\right)}{\left(2+k_-+k_+\right) \left(-4-4 k_--k_-^2-4
k_++3 k_- k_++4 k_-^2 k_+-k_+^2+4 k_- k_+^2+3 k_-^2 k_+^2\right)} \ ,\\
 w_{128}&=\tfrac{64 \left(-1+k_-\right) \left(1+k_-\right) \left(1+k_+\right) \left(2+2 k_-+k_+\right)}{\left(2+k_-+k_+\right) \left(-4-4 k_--k_-^2-4
k_++3 k_- k_++4 k_-^2 k_+-k_+^2+4 k_- k_+^2+3 k_-^2 k_+^2\right)} \ ,\\
 w_{132}&=\tfrac{64 \left(-1+k_-\right) \left(1+k_-\right) \left(2+2 k_-+k_+\right)}{\left(2+k_-+k_+\right) \left(-4-4 k_--k_-^2-4 k_++3 k_- k_++4 k_-^2
k_+-k_+^2+4 k_- k_+^2+3 k_-^2 k_+^2\right)} \ ,\\
 w_{138}&=\tfrac{8 \left(-1+k_-\right) \left(1+k_-\right) k_+ \left(2+2 k_-+k_+\right)}{\left(2+k_-+k_+\right) \left(-4-4 k_--k_-^2-4 k_++3 k_- k_++4
k_-^2 k_+-k_+^2+4 k_- k_+^2+3 k_-^2 k_+^2\right)} \ ,\\
 w_{143}&=\tfrac{16 k_- \left(1+k_-\right) \left(-1+k_+\right) \left(1+k_+\right) \left(2+k_-+2 k_+\right)}{\left(2+k_-+k_+\right) \left(-4-4 k_--k_-^2-4
k_++3 k_- k_++4 k_-^2 k_+-k_+^2+4 k_- k_+^2+3 k_-^2 k_+^2\right)} \ ,\\
 w_{144}&=-\tfrac{64 \left(1+k_-\right) \left(-1+k_+\right) \left(1+k_+\right) \left(2+k_-+2 k_+\right)}{\left(2+k_-+k_+\right) \left(-4-4 k_--k_-^2-4
k_++3 k_- k_++4 k_-^2 k_+-k_+^2+4 k_- k_+^2+3 k_-^2 k_+^2\right)} \ ,\\
 w_{152}&=-\tfrac{8 \left(1+k_-\right) \left(-1+k_+\right) \left(2+2 k_-+k_+\right) \left(2+k_-+2 k_+\right)}{\left(2+k_-+k_+\right) \left(-4-4 k_--k_-^2-4
k_++3 k_- k_++4 k_-^2 k_+-k_+^2+4 k_- k_+^2+3 k_-^2 k_+^2\right)} \ ,\\
 w_{159}&=\tfrac{16 k_- \left(1+k_-\right) \left(-1+k_+\right) \left(1+k_+\right) \left(2+k_-+2 k_+\right)}{\left(2+k_-+k_+\right) \left(-4-4 k_--k_-^2-4
k_++3 k_- k_++4 k_-^2 k_+-k_+^2+4 k_- k_+^2+3 k_-^2 k_+^2\right)} \ ,\\
 w_{160}&=-\tfrac{64 \left(1+k_-\right) \left(-1+k_+\right) \left(1+k_+\right) \left(2+k_-+2 k_+\right)}{\left(2+k_-+k_+\right) \left(-4-4 k_--k_-^2-4
k_++3 k_- k_++4 k_-^2 k_+-k_+^2+4 k_- k_+^2+3 k_-^2 k_+^2\right)} \ ,\\
 w_{164}&=-\tfrac{64 \left(-1+k_+\right) \left(1+k_+\right) \left(2+k_-+2 k_+\right)}{\left(2+k_-+k_+\right) \left(-4-4 k_--k_-^2-4 k_++3 k_- k_++4
k_-^2 k_+-k_+^2+4 k_- k_+^2+3 k_-^2 k_+^2\right)} \ ,\\
 w_{170}&=-\tfrac{8 k_- \left(-1+k_+\right) \left(1+k_+\right) \left(2+k_-+2 k_+\right)}{\left(2+k_-+k_+\right) \left(-4-4 k_--k_-^2-4 k_++3 k_- k_++4
k_-^2 k_+-k_+^2+4 k_- k_+^2+3 k_-^2 k_+^2\right)} \ ,\\
 w_{174}&=-\tfrac{32 k_- k_+ \left(1+k_-+2 k_+\right)}{\left(2+k_-+k_+\right){}^2} \ ,\\
 w_{175}&=-\tfrac{64 k_- \left(2+k_-\right) \left(-1+k_+\right) \left(1+k_+\right) \left(2+k_-+2 k_+\right)}{\left(2+k_-+k_+\right) \left(-4-4 k_--k_-^2-4
k_++3 k_- k_++4 k_-^2 k_+-k_+^2+4 k_- k_+^2+3 k_-^2 k_+^2\right)} \ ,\\
 w_{176}&=8 \ ,\\
 w_{177}&=\tfrac{16 \left(k_--k_+\right) \left(5+4 k_-+4 k_++2 k_- k_+\right)}{-4-4 k_--k_-^2-4 k_++3 k_- k_++4 k_-^2 k_+-k_+^2+4 k_- k_+^2+3 k_-^2
k_+^2} \ ,\\
 w_{178}&=-\tfrac{32 k_- k_+ \left(2+k_-+2 k_+\right) \left(-1+2 k_-+2 k_-^2-2 k_+-k_- k_+\right)}{\left(2+k_-+k_+\right){}^2 \left(-4-4 k_--k_-^2-4
k_++3 k_- k_++4 k_-^2 k_+-k_+^2+4 k_- k_+^2+3 k_-^2 k_+^2\right)} \ ,\\
 w_{179}&=\tfrac{32 k_- \left(-1+k_+\right) \left(1+k_+\right) \left(2+k_-+2 k_+\right)}{\left(2+k_-+k_+\right) \left(-4-4 k_--k_-^2-4 k_++3 k_- k_++4
k_-^2 k_+-k_+^2+4 k_- k_+^2+3 k_-^2 k_+^2\right)} \ ,\\
 w_{180}&=\tfrac{32 k_-}{\left(2+k_-+k_+\right){}^2} \ ,\\
 w_{181}&=-\tfrac{32 k_- \left(1+k_-+2 k_+\right)}{\left(2+k_-+k_+\right){}^2} \ ,\\
 w_{182}&=1 \ ,\\
 w_{183}&=\tfrac{4 \left(k_--k_+\right) \left(5+4 k_-+4 k_++2 k_- k_+\right)}{-4-4 k_--k_-^2-4 k_++3 k_- k_++4 k_-^2 k_+-k_+^2+4 k_- k_+^2+3 k_-^2 k_+^2}
\ ,\\
 w_{187}&=-\tfrac{32 \left(2+k_-+2 k_+\right) \left(-2-k_--k_++2 k_- k_++2 k_-^2 k_+\right)}{\left(2+k_-+k_+\right) \left(-4-4 k_--k_-^2-4 k_++3 k_-
k_++4 k_-^2 k_+-k_+^2+4 k_- k_+^2+3 k_-^2 k_+^2\right)} \ ,\\
 w_{188}&=\tfrac{2 \left(2+k_+\right) \left(2+k_-+2 k_+\right) \left(1-2 k_--2 k_-^2+2 k_++k_- k_+\right)}{\left(2+k_-+k_+\right) \left(-4-4 k_--k_-^2-4
k_++3 k_- k_++4 k_-^2 k_+-k_+^2+4 k_- k_+^2+3 k_-^2 k_+^2\right)} \ ,\\
 w_{190}&=\tfrac{2 \left(2+k_-+2 k_+\right) \left(-2-4 k_--k_++2 k_- k_++2 k_-^2 k_++3 k_- k_+^2\right)}{\left(2+k_-+k_+\right) \left(-4-4 k_--k_-^2-4
k_++3 k_- k_++4 k_-^2 k_+-k_+^2+4 k_- k_+^2+3 k_-^2 k_+^2\right)} \ ,\\
 w_{192}&=\tfrac{8 k_- \left(-4-4 k_--k_-^2-8 k_+-k_- k_++8 k_-^2 k_++4 k_-^3 k_+-3 k_+^2+4 k_- k_+^2+5 k_-^2 k_+^2\right)}{\left(2+k_-+k_+\right){}^2
\left(-4-4 k_--k_-^2-4 k_++3 k_- k_++4 k_-^2 k_+-k_+^2+4 k_- k_+^2+3 k_-^2 k_+^2\right)} \ ,\\
 w_{193}&=\tfrac{16 \left(2+k_-+2 k_+\right) \left(-2-k_--k_++2 k_- k_++2 k_-^2 k_+\right)}{\left(2+k_-+k_+\right){}^2 \left(-4-4 k_--k_-^2-4 k_++3
k_- k_++4 k_-^2 k_+-k_+^2+4 k_- k_+^2+3 k_-^2 k_+^2\right)} \ ,\\
 w_{194}&=\tfrac{16 \left(2+k_-+2 k_+\right) \left(-2-3 k_--k_++2 k_- k_++2 k_-^2 k_++2 k_- k_+^2\right)}{\left(2+k_-+k_+\right){}^2 \left(-4-4 k_--k_-^2-4
k_++3 k_- k_++4 k_-^2 k_+-k_+^2+4 k_- k_+^2+3 k_-^2 k_+^2\right)} \ ,\\
 w_{195}&=-\tfrac{32 k_- k_+ \left(1+2 k_-+k_+\right)}{\left(2+k_-+k_+\right){}^2} \ ,\\
 w_{196}&=-\tfrac{64 \left(-1+k_-\right) \left(1+k_-\right) k_+ \left(2+k_+\right) \left(2+2 k_-+k_+\right)}{\left(2+k_-+k_+\right) \left(-4-4 k_--k_-^2-4
k_++3 k_- k_++4 k_-^2 k_+-k_+^2+4 k_- k_+^2+3 k_-^2 k_+^2\right)} \ ,\\
 w_{197}&=-8 \ ,\\
 w_{198}&=-\tfrac{16 \left(k_--k_+\right) \left(5+4 k_-+4 k_++2 k_- k_+\right)}{-4-4 k_--k_-^2-4 k_++3 k_- k_++4 k_-^2 k_+-k_+^2+4 k_- k_+^2+3 k_-^2
k_+^2} \ ,\\
 w_{199}&=\tfrac{32 \left(-1+k_-\right) \left(1+k_-\right) k_+ \left(2+2 k_-+k_+\right)}{\left(2+k_-+k_+\right) \left(-4-4 k_--k_-^2-4 k_++3 k_- k_++4
k_-^2 k_+-k_+^2+4 k_- k_+^2+3 k_-^2 k_+^2\right)} \ ,\\
 w_{200}&=\tfrac{32 k_- k_+ \left(2+2 k_-+k_+\right) \left(1+2 k_--2 k_++k_- k_+-2 k_+^2\right)}{\left(2+k_-+k_+\right){}^2 \left(-4-4 k_--k_-^2-4 k_++3
k_- k_++4 k_-^2 k_+-k_+^2+4 k_- k_+^2+3 k_-^2 k_+^2\right)} \ ,\\
 w_{201}&=\tfrac{32 k_+}{\left(2+k_-+k_+\right){}^2} \ ,\\
 w_{202}&=-\tfrac{32 k_+ \left(1+2 k_-+k_+\right)}{\left(2+k_-+k_+\right){}^2} \ ,\\
 w_{203}&=-1 \ ,\\
 w_{204}&=-\tfrac{4 \left(k_--k_+\right) \left(5+4 k_-+4 k_++2 k_- k_+\right)}{-4-4 k_--k_-^2-4 k_++3 k_- k_++4 k_-^2 k_+-k_+^2+4 k_- k_+^2+3 k_-^2
k_+^2} \ ,\\
 w_{208}&=-\tfrac{32 \left(2+2 k_-+k_+\right) \left(-2-k_--k_++2 k_- k_++2 k_- k_+^2\right)}{\left(2+k_-+k_+\right) \left(-4-4 k_--k_-^2-4 k_++3 k_-
k_++4 k_-^2 k_+-k_+^2+4 k_- k_+^2+3 k_-^2 k_+^2\right)} \ ,\\
 w_{209}&=\tfrac{2 \left(2+k_-\right) \left(2+2 k_-+k_+\right) \left(1+2 k_--2 k_++k_- k_+-2 k_+^2\right)}{\left(2+k_-+k_+\right) \left(-4-4 k_--k_-^2-4
k_++3 k_- k_++4 k_-^2 k_+-k_+^2+4 k_- k_+^2+3 k_-^2 k_+^2\right)} \ ,\\
 w_{211}&=\tfrac{2 \left(2+2 k_-+k_+\right) \left(-2-k_--4 k_++2 k_- k_++3 k_-^2 k_++2 k_- k_+^2\right)}{\left(2+k_-+k_+\right) \left(-4-4 k_--k_-^2-4
k_++3 k_- k_++4 k_-^2 k_+-k_+^2+4 k_- k_+^2+3 k_-^2 k_+^2\right)} \ ,\\
 w_{213}&=\tfrac{8 k_+ \left(-4-8 k_--3 k_-^2-4 k_+-k_- k_++4 k_-^2 k_+-k_+^2+8 k_- k_+^2+5 k_-^2 k_+^2+4 k_- k_+^3\right)}{\left(2+k_-+k_+\right){}^2
\left(-4-4 k_--k_-^2-4 k_++3 k_- k_++4 k_-^2 k_+-k_+^2+4 k_- k_+^2+3 k_-^2 k_+^2\right)} \ ,\\
 w_{214}&=\tfrac{16 \left(2+2 k_-+k_+\right) \left(-2-k_--k_++2 k_- k_++2 k_- k_+^2\right)}{\left(2+k_-+k_+\right){}^2 \left(-4-4 k_--k_-^2-4 k_++3
k_- k_++4 k_-^2 k_+-k_+^2+4 k_- k_+^2+3 k_-^2 k_+^2\right)} \ ,\\
 w_{215}&=\tfrac{16 \left(2+2 k_-+k_+\right) \left(-2-k_--3 k_++2 k_- k_++2 k_-^2 k_++2 k_- k_+^2\right)}{\left(2+k_-+k_+\right){}^2 \left(-4-4 k_--k_-^2-4
k_++3 k_- k_++4 k_-^2 k_+-k_+^2+4 k_- k_+^2+3 k_-^2 k_+^2\right)} \ ,\\
 w_{216}&=-\tfrac{32}{2+k_-+k_+} \ ,\\
 w_{221}&=-\tfrac{32}{2+k_-+k_+} \ ,\\
 w_{223}&=-\tfrac{64 \left(k_--k_+\right) \left(8+8 k_-+2 k_-^2+8 k_++9 k_- k_++4 k_-^2 k_++2 k_+^2+4 k_- k_+^2\right)}{3 \left(2+k_-+k_+\right) \left(-4-4
k_--k_-^2-4 k_++3 k_- k_++4 k_-^2 k_+-k_+^2+4 k_- k_+^2+3 k_-^2 k_+^2\right)} \ ,\\
 w_{224}&=-8 \ ,\\
 w_{225}&=-\tfrac{16 \left(k_--k_+\right) \left(5+4 k_-+4 k_++2 k_- k_+\right)}{-4-4 k_--k_-^2-4 k_++3 k_- k_++4 k_-^2 k_+-k_+^2+4 k_- k_+^2+3 k_-^2
k_+^2} \ ,\\
w_{226}&=\tfrac{32 \left[3 k_- k_+ \left(2+k_-+2 k_+\right) \left(-1+2 k_-+2 k_-^2-2 k_+-k_- k_+\right)+\left(k_-+2k_+\right)K\right]}{3
\left(2+k_-+k_+\right){}^2 K}\ ,\\
w_{227}&=\tfrac{32 \left[3 k_- k_+ \left(2+2 k_-+k_+\right) \left(1+2 k_--2 k_++k_- k_+-2 k_+^2\right)-\left(2k_-+k_+\right)K\right]}{3
\left(2+k_-+k_+\right){}^2 K}\ ,\\
 w_{238}&=-\tfrac{128 k_- \left(3+k_-+2 k_+\right)}{3 \left(2+k_-+k_+\right){}^2} \ ,\\
 w_{241}&=1 \ ,\\
 w_{242}&=\tfrac{4 \left(k_--k_+\right) \left(5+4 k_-+4 k_++2 k_- k_+\right)}{-4-4 k_--k_-^2-4 k_++3 k_- k_++4 k_-^2 k_+-k_+^2+4 k_- k_+^2+3 k_-^2 k_+^2}
\ ,\\
 w_{246}&=-\tfrac{32 \left(2+k_-+2 k_+\right) \left(-2-k_--k_++2 k_- k_++2 k_-^2 k_+\right)}{\left(2+k_-+k_+\right) \left(-4-4 k_--k_-^2-4 k_++3 k_-
k_++4 k_-^2 k_+-k_+^2+4 k_- k_+^2+3 k_-^2 k_+^2\right)} \ ,\\
 w_{247}&=-\tfrac{2 \left(8+18 k_-+15 k_-^2+4 k_-^3-2 k_+-6 k_- k_++2 k_-^3 k_+-11 k_+^2-16 k_- k_+^2-6 k_-^2 k_+^2-4 k_+^3-2 k_- k_+^3\right)}{\left(2+k_-+k_+\right)
\left(-4-4 k_--k_-^2-4 k_++3 k_- k_++4 k_-^2 k_+-k_+^2+4 k_- k_+^2+3 k_-^2 k_+^2\right)} \ ,\\
 w_{249}&=\tfrac{2 \left(-8-14 k_--5 k_-^2-10 k_+-2 k_- k_++10 k_-^2 k_++2 k_-^3 k_+-3 k_+^2+14 k_- k_+^2+10 k_-^2 k_+^2+6 k_- k_+^3\right)}{\left(2+k_-+k_+\right)
\left(-4-4 k_--k_-^2-4 k_++3 k_- k_++4 k_-^2 k_+-k_+^2+4 k_- k_+^2+3 k_-^2 k_+^2\right)} \ ,\\
 w_{251}&=\tfrac{16 k_- \left(2+k_-+2 k_+\right) \left(-2-k_--k_++2 k_- k_++2 k_-^2 k_+\right)}{\left(2+k_-+k_+\right){}^2 \left(-4-4 k_--k_-^2-4 k_++3
k_- k_++4 k_-^2 k_+-k_+^2+4 k_- k_+^2+3 k_-^2 k_+^2\right)} \ ,\\
 w_{252}&=\tfrac{16 \left(2+k_-+2 k_+\right) \left(-2-k_--k_++2 k_- k_++2 k_-^2 k_+\right)}{\left(2+k_-+k_+\right){}^2 \left(-4-4 k_--k_-^2-4 k_++3
k_- k_++4 k_-^2 k_+-k_+^2+4 k_- k_+^2+3 k_-^2 k_+^2\right)} \ ,\\
 w_{253}&=\tfrac{16 \left(-12-16 k_--5 k_-^2-14 k_++3 k_- k_++14 k_-^2 k_++2 k_-^3 k_+-4 k_+^2+16 k_- k_+^2+12 k_-^2 k_+^2+4 k_- k_+^3\right)}{\left(2+k_-+k_+\right){}^2
\left(-4-4 k_--k_-^2-4 k_++3 k_- k_++4 k_-^2 k_+-k_+^2+4 k_- k_+^2+3 k_-^2 k_+^2\right)} \ ,\\
 w_{254}&=\tfrac{128 k_+ \left(3+2 k_-+k_+\right)}{3 \left(2+k_-+k_+\right){}^2} \ ,\\
 w_{257}&=1 \ ,\\
 w_{258}&=\tfrac{4 \left(k_--k_+\right) \left(5+4 k_-+4 k_++2 k_- k_+\right)}{-4-4 k_--k_-^2-4 k_++3 k_- k_++4 k_-^2 k_+-k_+^2+4 k_- k_+^2+3 k_-^2 k_+^2}
\ ,\\
 w_{262}&=\tfrac{32 \left(2+2 k_-+k_+\right) \left(-2-k_--k_++2 k_- k_++2 k_- k_+^2\right)}{\left(2+k_-+k_+\right) \left(-4-4 k_--k_-^2-4 k_++3 k_-
k_++4 k_-^2 k_+-k_+^2+4 k_- k_+^2+3 k_-^2 k_+^2\right)} \ ,\\
 w_{263}&=-\tfrac{2 \left(-8+2 k_-+11 k_-^2+4 k_-^3-18 k_++6 k_- k_++16 k_-^2 k_++2 k_-^3 k_+-15 k_+^2+6 k_-^2 k_+^2-4 k_+^3-2 k_- k_+^3\right)}{\left(2+k_-+k_+\right)
\left(-4-4 k_--k_-^2-4 k_++3 k_- k_++4 k_-^2 k_+-k_+^2+4 k_- k_+^2+3 k_-^2 k_+^2\right)} \ ,\\
 w_{265}&=-\tfrac{2 \left(-8-10 k_--3 k_-^2-14 k_+-2 k_- k_++14 k_-^2 k_++6 k_-^3 k_+-5 k_+^2+10 k_- k_+^2+10 k_-^2 k_+^2+2 k_- k_+^3\right)}{\left(2+k_-+k_+\right)
\left(-4-4 k_--k_-^2-4 k_++3 k_- k_++4 k_-^2 k_+-k_+^2+4 k_- k_+^2+3 k_-^2 k_+^2\right)} \ ,\\
 w_{267}&=-\tfrac{16 k_+ \left(2+2 k_-+k_+\right) \left(-2-k_--k_++2 k_- k_++2 k_- k_+^2\right)}{\left(2+k_-+k_+\right){}^2 \left(-4-4 k_--k_-^2-4 k_++3
k_- k_++4 k_-^2 k_+-k_+^2+4 k_- k_+^2+3 k_-^2 k_+^2\right)} \ ,\\
 w_{268}&=-\tfrac{16 \left(2+2 k_-+k_+\right) \left(-2-k_--k_++2 k_- k_++2 k_- k_+^2\right)}{\left(2+k_-+k_+\right){}^2 \left(-4-4 k_--k_-^2-4 k_++3
k_- k_++4 k_-^2 k_+-k_+^2+4 k_- k_+^2+3 k_-^2 k_+^2\right)} \ ,\\
 w_{269}&=-\tfrac{16 \left(-12-14 k_--4 k_-^2-16 k_++3 k_- k_++16 k_-^2 k_++4 k_-^3 k_+-5 k_+^2+14 k_- k_+^2+12 k_-^2 k_+^2+2 k_- k_+^3\right)}{\left(2+k_-+k_+\right){}^2
\left(-4-4 k_--k_-^2-4 k_++3 k_- k_++4 k_-^2 k_+-k_+^2+4 k_- k_+^2+3 k_-^2 k_+^2\right)} \ ,\\
 w_{270}&=-\tfrac{128 \left(k_--k_+\right)}{3 \left(2+k_-+k_+\right){}^2} \ ,\\
 w_{275}&=-\tfrac{64}{2+k_-+k_+} \ ,\\
 w_{284}&=-1 \ ,\\
 w_{286}&=\tfrac{16 \left(k_--k_+\right) \left(5+4 k_-+4 k_++2 k_- k_+\right)}{-4-4 k_--k_-^2-4 k_++3 k_- k_++4 k_-^2 k_+-k_+^2+4 k_- k_+^2+3 k_-^2
k_+^2} \ ,\\
 w_{287}&=\tfrac{2 \left(k_--k_+\right) \left(8+8 k_-+2 k_-^2+8 k_++9 k_- k_++4 k_-^2 k_++2 k_+^2+4 k_- k_+^2\right)}{\left(2+k_-+k_+\right) \left(-4-4
k_--k_-^2-4 k_++3 k_- k_++4 k_-^2 k_+-k_+^2+4 k_- k_+^2+3 k_-^2 k_+^2\right)} \ ,\\
 w_{288}&=\tfrac{4 \left(2+k_-+2 k_+\right) \left(-2-k_--k_++2 k_- k_++2 k_-^2 k_+\right)}{\left(2+k_-+k_+\right) \left(-4-4 k_--k_-^2-4 k_++3 k_- k_++4
k_-^2 k_+-k_+^2+4 k_- k_+^2+3 k_-^2 k_+^2\right)} \ ,\\
 w_{289}&=-\tfrac{4 \left(2+2 k_-+k_+\right) \left(-2-k_--k_++2 k_- k_++2 k_- k_+^2\right)}{\left(2+k_-+k_+\right) \left(-4-4 k_--k_-^2-4 k_++3 k_-
k_++4 k_-^2 k_+-k_+^2+4 k_- k_+^2+3 k_-^2 k_+^2\right)} \ ,\\
 w_{290}&=-\tfrac{8 \left(k_--k_+\right) \left(5+4 k_-+4 k_++2 k_- k_+\right)}{\left(2+k_-+k_+\right) \left(-4-4 k_--k_-^2-4 k_++3 k_- k_++4 k_-^2 k_+-k_+^2+4
k_- k_+^2+3 k_-^2 k_+^2\right)} \ ,\\
 w_{291}&=-\tfrac{8 \left(k_--k_+\right) \left(5+4 k_-+4 k_++2 k_- k_+\right)}{\left(2+k_-+k_+\right) \left(-4-4 k_--k_-^2-4 k_++3 k_- k_++4 k_-^2 k_+-k_+^2+4
k_- k_+^2+3 k_-^2 k_+^2\right)} \ ,\\
 w_{293}&=\tfrac{16 \left(-1+k_-\right) \left(1+k_-\right) k_+ \left(1+k_+\right) \left(2+2 k_-+k_+\right)}{\left(2+k_-+k_+\right) \left(-4-4 k_--k_-^2-4
k_++3 k_- k_++4 k_-^2 k_+-k_+^2+4 k_- k_+^2+3 k_-^2 k_+^2\right)} \ ,\\
 w_{294}&=-\tfrac{64 \left(-1+k_-\right) \left(1+k_-\right) \left(1+k_+\right) \left(2+2 k_-+k_+\right)}{\left(2+k_-+k_+\right) \left(-4-4 k_--k_-^2-4
k_++3 k_- k_++4 k_-^2 k_+-k_+^2+4 k_- k_+^2+3 k_-^2 k_+^2\right)} \ ,\\
 w_{298}&=-\tfrac{32 \left(-1+k_-\right) \left(1+k_-\right) \left(2+2 k_-+k_+\right)}{\left(2+k_-+k_+\right) \left(-4-4 k_--k_-^2-4 k_++3 k_- k_++4
k_-^2 k_+-k_+^2+4 k_- k_+^2+3 k_-^2 k_+^2\right)} \ ,\\
 w_{302}&=-\tfrac{4 \left(-1+k_-\right) \left(1+k_+\right) \left(2+2 k_-+k_+\right) \left(2+k_-+2 k_+\right)}{\left(2+k_-+k_+\right) \left(-4-4 k_--k_-^2-4
k_++3 k_- k_++4 k_-^2 k_+-k_+^2+4 k_- k_+^2+3 k_-^2 k_+^2\right)} \ ,\\
 w_{304}&=-\tfrac{4 \left(-1+k_-\right) \left(1+k_-\right) k_+ \left(2+2 k_-+k_+\right)}{\left(2+k_-+k_+\right) \left(-4-4 k_--k_-^2-4 k_++3 k_- k_++4
k_-^2 k_+-k_+^2+4 k_- k_+^2+3 k_-^2 k_+^2\right)} \ ,\\
 w_{309}&=-\tfrac{16 k_- \left(1+k_-\right) \left(-1+k_+\right) \left(1+k_+\right) \left(2+k_-+2 k_+\right)}{\left(2+k_-+k_+\right) \left(-4-4 k_--k_-^2-4
k_++3 k_- k_++4 k_-^2 k_+-k_+^2+4 k_- k_+^2+3 k_-^2 k_+^2\right)} \ ,\\
 w_{310}&=\tfrac{64 \left(1+k_-\right) \left(-1+k_+\right) \left(1+k_+\right) \left(2+k_-+2 k_+\right)}{\left(2+k_-+k_+\right) \left(-4-4 k_--k_-^2-4
k_++3 k_- k_++4 k_-^2 k_+-k_+^2+4 k_- k_+^2+3 k_-^2 k_+^2\right)} \ ,\\
 w_{314}&=\tfrac{32 \left(-1+k_+\right) \left(1+k_+\right) \left(2+k_-+2 k_+\right)}{\left(2+k_-+k_+\right) \left(-4-4 k_--k_-^2-4 k_++3 k_- k_++4 k_-^2
k_+-k_+^2+4 k_- k_+^2+3 k_-^2 k_+^2\right)} \ ,\\
 w_{318}&=\tfrac{4 \left(1+k_-\right) \left(-1+k_+\right) \left(2+2 k_-+k_+\right) \left(2+k_-+2 k_+\right)}{\left(2+k_-+k_+\right) \left(-4-4 k_--k_-^2-4
k_++3 k_- k_++4 k_-^2 k_+-k_+^2+4 k_- k_+^2+3 k_-^2 k_+^2\right)} \ ,\\
 w_{320}&=\tfrac{4 k_- \left(-1+k_+\right) \left(1+k_+\right) \left(2+k_-+2 k_+\right)}{\left(2+k_-+k_+\right) \left(-4-4 k_--k_-^2-4 k_++3 k_- k_++4
k_-^2 k_+-k_+^2+4 k_- k_+^2+3 k_-^2 k_+^2\right)} \ ,\\
 w_{325}&=\tfrac{8 \left(k_--k_+\right) \left(2+2 k_-+k_+\right) \left(2+k_-+2 k_+\right)}{\left(2+k_-+k_+\right) \left(-4-4 k_--k_-^2-4 k_++3 k_- k_++4
k_-^2 k_+-k_+^2+4 k_- k_+^2+3 k_-^2 k_+^2\right)} \ ,\\
 w_{326}&=\tfrac{8 \left(2+k_-+2 k_+\right) \left(-2-k_--k_++2 k_- k_++2 k_-^2 k_+\right)}{\left(2+k_-+k_+\right) \left(-4-4 k_--k_-^2-4 k_++3 k_- k_++4
k_-^2 k_+-k_+^2+4 k_- k_+^2+3 k_-^2 k_+^2\right)} \ ,\\
 w_{327}&=-\tfrac{8 \left(2+2 k_-+k_+\right) \left(-2-k_--k_++2 k_- k_++2 k_- k_+^2\right)}{\left(2+k_-+k_+\right) \left(-4-4 k_--k_-^2-4 k_++3 k_-
k_++4 k_-^2 k_+-k_+^2+4 k_- k_+^2+3 k_-^2 k_+^2\right)} \ ,\\
 w_{328}&=\tfrac{32 \left(k_--k_+\right) \left(5+4 k_-+4 k_++2 k_- k_+\right)}{\left(2+k_-+k_+\right) \left(-4-4 k_--k_-^2-4 k_++3 k_- k_++4 k_-^2 k_+-k_+^2+4
k_- k_+^2+3 k_-^2 k_+^2\right)} \ ,\\
 n_2 &=\tfrac{64 \left(-1+k_-\right) k_- \left(1+k_-\right) \left(-1+k_+\right) k_+ \left(1+k_+\right) \left(2+2 k_-+k_+\right) \left(2+k_-+2 k_+\right)}{\left(2+k_-+k_+\right){}^3
\left(-4-4 k_--k_-^2-4 k_++3 k_- k_++4 k_-^2 k_+-k_+^2+4 k_- k_+^2+3 k_-^2 k_+^2\right)} \ ,\\
 w_{330}&=\tfrac{128 \left(-1+k_-\right) k_- \left(1+k_-\right) \left(2+k_-\right) \left(-1+k_+\right) k_+ \left(1+k_+\right) \left(2+k_+\right) \left(2+2
k_-+k_+\right) \left(2+k_-+2 k_+\right)}{\left(2+k_-+k_+\right){}^2 \left(-4-4 k_--k_-^2-4 k_++3 k_- k_++4 k_-^2 k_+-k_+^2+4 k_- k_+^2+3 k_-^2 k_+^2\right){}^2}
\ ,\\
 w_{331}&=-\tfrac{16 \left(k_--k_+\right) \left(4+4 k_-+k_-^2+4 k_++7 k_- k_++4 k_-^2 k_++k_+^2+4 k_- k_+^2+k_-^2 k_+^2\right)}{\left(2+k_-+k_+\right)
\left(-4-4 k_--k_-^2-4 k_++3 k_- k_++4 k_-^2 k_+-k_+^2+4 k_- k_+^2+3 k_-^2 k_+^2\right)} \ ,\\
 w_{332}&=\tfrac{32 \left(-1+k_-\right) \left(1+k_-\right) \left(2+k_-\right) \left(-1+k_+\right) \left(1+k_+\right) \left(2+k_+\right) \left(2+2 k_-+k_+\right)
\left(2+k_-+2 k_+\right)}{\left(2+k_-+k_+\right) \left(-4-4 k_--k_-^2-4 k_++3 k_- k_++4 k_-^2 k_+-k_+^2+4 k_- k_+^2+3 k_-^2 k_+^2\right){}^2} \ ,\\
 w_{333}&=-\tfrac{64 \left(-1+k_-\right) k_- \left(1+k_-\right) \left(2+k_-\right) \left(-1+k_+\right) k_+ \left(1+k_+\right) \left(2+2 k_-+k_+\right)
\left(2+k_-+2 k_+\right)}{\left(2+k_-+k_+\right){}^2 \left(-4-4 k_--k_-^2-4 k_++3 k_- k_++4 k_-^2 k_+-k_+^2+4 k_- k_+^2+3 k_-^2 k_+^2\right){}^2}
\ ,\\
 w_{334}&=-\tfrac{64 \left(-1+k_-\right) k_- \left(1+k_-\right) \left(-1+k_+\right) k_+ \left(1+k_+\right) \left(2+k_+\right) \left(2+2 k_-+k_+\right)
\left(2+k_-+2 k_+\right)}{\left(2+k_-+k_+\right){}^2 \left(-4-4 k_--k_-^2-4 k_++3 k_- k_++4 k_-^2 k_+-k_+^2+4 k_- k_+^2+3 k_-^2 k_+^2\right){}^2}\ .
\end{align*}

\section{Structure Constants for the OPE  $V^{(1)}\times V^{(1)}$}
\label{app:1x1-linear}

In this appendix we list the values of the $39$ coefficients appearing in (\ref{eq:110-solution}).

\begin{align*}
z_1&= -\frac{\left(k_-+k_+\right) n_1}{2 \left(k_--1\right)
   \left(k_+-1\right)}\ ,&
z_2&= -\frac{n_1}{\left(k_--1\right)
   \left(k_+-1\right)}\ ,\\
z_3&= \frac{2 n_1}{\left(k_--1\right)
   \left(k_+-1\right)}\ ,&
z_4&= -\frac{2 n_1}{\left(k_--1\right)
   \left(k_+-1\right)}\ ,\\
z_5&= -\frac{2 n_1}{3 \left(k_--1\right)
   \left(k_+-1\right) \left(k_-+k_+\right)}\ ,&
z_6&= -2 n_1\ ,\\
z_7&= \frac{4   n_1}{k_--1}\ ,&
z_8&= \frac{4 n_1}{k_+-1}\ ,\\
z_9&= -\frac{2    \left(k_-+k_+-2\right) n_1}{\left(k_--1\right) \left(k_+-1\right)
   \left(k_-+k_+\right)},&
z_{10}&= \frac{\left(k_+-k_-\right)
   n_1}{\left(k_--1\right) \left(k_+-1\right) \left(k_-+k_+\right)}\ ,\\
z_{11}&=
   -\frac{n_1}{\left(k_--1\right) \left(k_+-1\right)}\ ,&
z_{12}&= -\frac{8
   n_1}{\left(k_--1\right) \left(k_+-1\right)}\ ,\\
z_{13}&=
   -\frac{n_1}{\left(k_--1\right) \left(k_+-1\right)}\ ,&
z_{14}&= \frac{2
   n_1}{k_--1}\ ,\\
z_{15}&= \frac{4 n_1}{\left(k_--1\right) \left(k_+-1\right)
   \left(k_-+k_+\right)}\ ,&
z_{16}&= \frac{2 n_1}{\left(k_--1\right)
   \left(k_+-1\right) \left(k_-+k_+\right)}\ ,\\
z_{17}&= \frac{2
   n_1}{\left(k_--1\right) \left(k_+-1\right) \left(k_-+k_+\right)}\ ,&
z_{18}&=
   \frac{4 n_1}{\left(k_--1\right) \left(k_+-1\right)
   \left(k_-+k_+\right)}\ ,\\
z_{19}&= -\frac{2 n_1}{\left(k_--1\right)
   \left(k_+-1\right) \left(k_-+k_+\right)}\ ,&
z_{20}&= \frac{2
   n_1}{\left(k_--1\right) \left(k_+-1\right) \left(k_-+k_+\right)}\ ,\\
z_{21}&=
   \frac{2 n_1}{k_+-1}\ ,&
z_{22}&= -\frac{n_1}{\left(k_--1\right)
   \left(k_+-1\right)}\ ,\\
z_{23}&= -\frac{\left(k_-+k_+-4\right)
   n_1}{\left(k_--1\right) \left(k_+-1\right) \left(k_-+k_+\right)}\ ,&
z_{24}&=
   \frac{\left(k_-+k_++4\right) n_1}{\left(k_--1\right) \left(k_+-1\right)
   \left(k_-+k_+\right)}\ ,\\
z_{25}&= \frac{\left(k_+-k_-\right)
   n_1}{\left(k_--1\right) \left(k_+-1\right) \left(k_-+k_+\right)}\ ,&
z_{26}&=
   -\frac{\left(k_-+k_+\right) n_1}{\left(k_--1\right)
   \left(k_+-1\right)}\ ,\\
z_{27}&= -\frac{n_1}{\left(k_--1\right)
   \left(k_+-1\right)}\ ,&
z_{28}&= -\frac{4 n_1}{k_+-1}\ ,\\
z_{29}&= \frac{4
   n_1}{\left(k_+-1\right) \left(k_-+k_+\right)}\ ,&
z_{30}&= -\frac{4
   n_1}{\left(k_--1\right) \left(k_+-1\right) \left(k_-+k_+\right)}\ ,\\
z_{31}&=
   -\frac{8 n_1}{\left(k_--1\right) \left(k_+-1\right)
   \left(k_-+k_+\right)}\ ,&
z_{32}&= \frac{2 n_1}{\left(k_--1\right)
   \left(k_+-1\right)}\ ,\\
z_{33}&= \frac{4 n_1}{\left(k_--1\right)
   \left(k_+-1\right) \left(k_-+k_+\right)}\ ,&
z_{34}&= -\frac{4
   n_1}{k_--1}\ ,\\
z_{35}&= \frac{4 n_1}{\left(k_--1\right)
   \left(k_-+k_+\right)}\ ,&
z_{36}&= -\frac{4 n_1}{\left(k_--1\right)
   \left(k_+-1\right) \left(k_-+k_+\right)}\ ,\\
z_{37}&= -\frac{8
   n_1}{\left(k_--1\right) \left(k_+-1\right) \left(k_-+k_+\right)}\ ,&
z_{38}&=
   -\frac{2 n_1}{\left(k_--1\right) \left(k_+-1\right)}\ ,\\
z_{39}&= -\frac{4
   n_1}{\left(k_--1\right) \left(k_+-1\right) \left(k_-+k_+\right)} \ .
\end{align*}

\section{Structure Constants for the OPE $V^{(1)}\times V^{(2)}$}
\label{app:1x2-linear}

In this appendix we list the values of the 16 coefficients in eq.~(\ref{eq:Phi52}). We have omitted a 
common factor $n_{1}/n_{2}$, where $n_{1}$ is defined in (\ref{eq:110-solution}), and
$n_{2}$ is the coefficient of $V^{(2)}_{1/2}$ in the OPE $V^{(1)}_{0}\times V^{(1)}_{3/2}$.

\begin{align*}
w_1 &=  -\tfrac{k_-+k_+}{2 \left(k_--1\right) \left(k_+-1\right)}\ ,\\
w_2 &= 
   -\tfrac{2 \left(k_--k_+\right) \left(2 k_+ k_-^2-10 k_-^2+2 k_+^2 k_--26 k_+
   k_-+17 k_--10 k_+^2+17 k_+-6\right)}{3 \left(k_--1\right) \left(k_+-1\right)
   \left(3 k_+^2 k_-^2-2 k_+ k_-^2-2 k_-^2-2 k_+^2 k_--k_+ k_-+k_--2
   k_+^2+k_+\right)}\ ,\\
   w_3 &=  -\tfrac{4}{\left(k_--1\right)
   \left(k_+-1\right)}\ ,\\
   w_4 &=  \tfrac{2}{\left(k_--1\right)
   \left(k_+-1\right)}\ ,\\
   w_5 &=  -\tfrac{2 \left(k_--k_+\right) \left(k_-+k_+\right)
   \left(2 k_+ k_-+2 k_-+2 k_+-1\right)}{\left(k_--1\right) \left(k_+-1\right)
   \left(3 k_+^2 k_-^2-2 k_+ k_-^2-2 k_-^2-2 k_+^2 k_--k_+ k_-+k_--2
   k_+^2+k_+\right)}\ ,\\
   w_6 &=  \tfrac{2}{\left(k_--1\right) \left(k_+-1\right)}\ ,\\
   w_7 &= 
   -\tfrac{4 \left(k_--k_+\right) \left(2 k_+ k_-+2 k_-+2
   k_+-1\right)}{\left(k_--1\right) \left(k_+-1\right) \left(3 k_+^2 k_-^2-2 k_+
   k_-^2-2 k_-^2-2 k_+^2 k_--k_+ k_-+k_--2 k_+^2+k_+\right)}\ ,\\
   w_8 &=  -\tfrac{4
   \left(4 k_+ k_-^3-4 k_-^3+5 k_+^2 k_-^2-14 k_+ k_-^2+8 k_-^2-6 k_+^2 k_-+7 k_+
   k_--3 k_--2 k_+^2+k_+\right)}{\left(k_--1\right) \left(k_+-1\right) \left(3
   k_+^2 k_-^2-2 k_+ k_-^2-2 k_-^2-2 k_+^2 k_--k_+ k_-+k_--2
   k_+^2+k_+\right)}\ ,\\
   w_9 &=  \tfrac{4 \left(4 k_- k_+^3-4 k_+^3+5 k_-^2 k_+^2-14 k_-
   k_+^2+8 k_+^2-6 k_-^2 k_++7 k_- k_+-3 k_+-2
   k_-^2+k_-\right)}{\left(k_--1\right) \left(k_+-1\right) \left(3 k_+^2 k_-^2-2
   k_+ k_-^2-2 k_-^2-2 k_+^2 k_--k_+ k_-+k_--2 k_+^2+k_+\right)}\ ,\\
   w_{10} &=  \tfrac{8
   \left(k_--k_+\right) \left(2 k_+ k_-+2 k_-+2 k_+-1\right)}{\left(k_--1\right)
   \left(k_+-1\right) \left(3 k_+^2 k_-^2-2 k_+ k_-^2-2 k_-^2-2 k_+^2 k_--k_+
   k_-+k_--2 k_+^2+k_+\right)}\ ,\\
   w_{11} &=  -\tfrac{8 \left(k_--k_+\right) \left(2 k_+
   k_-+2 k_-+2 k_+-1\right)}{\left(k_--1\right) \left(k_+-1\right) \left(3 k_+^2
   k_-^2-2 k_+ k_-^2-2 k_-^2-2 k_+^2 k_--k_+ k_-+k_--2 k_+^2+k_+\right)}\ ,\\
   w_{12} &= 
   \tfrac{2}{\left(k_--1\right) \left(k_+-1\right)}\ ,\\
   w_{13} &= 
   \tfrac{2}{\left(k_--1\right) \left(k_+-1\right)}\ ,\\
   w_{14} &=  -\tfrac{8
   \left(k_--k_+\right) \left(2 k_+ k_-+2 k_-+2 k_+-1\right)}{3 \left(k_--1\right)
   \left(k_+-1\right) \left(k_-+k_+\right) \left(3 k_+^2 k_-^2-2 k_+ k_-^2-2
   k_-^2-2 k_+^2 k_--k_+ k_-+k_--2 k_+^2+k_+\right)}\ ,\nonumber \\
   &&\\
   w_{15} &=  \tfrac{8 \left(k_-+2
   k_+-1\right) \left(2 k_+ k_-^2-2 k_-^2-2 k_+
   k_-+k_--k_+\right)}{\left(k_--1\right) \left(k_+-1\right) \left(k_-+k_+\right)
   \left(3 k_+^2 k_-^2-2 k_+ k_-^2-2 k_-^2-2 k_+^2 k_--k_+ k_-+k_--2
   k_+^2+k_+\right)}\ ,\nonumber \\
   && \\
   w_{16} &=  -\tfrac{8 \left(2 k_-+k_+-1\right) \left(2 k_-
   k_+^2-2 k_+^2-2 k_- k_++k_+-k_-\right)}{\left(k_--1\right) \left(k_+-1\right)
   \left(k_-+k_+\right) \left(3 k_+^2 k_-^2-2 k_+ k_-^2-2 k_-^2-2 k_+^2 k_--k_+
   k_-+k_--2 k_+^2+k_+\right)}\ .\nonumber 
\end{align*}


\begin{thebibliography}{99}

\bibitem{Vasiliev:1995dn}
M.A.~Vasiliev,
``Higher spin gauge theories in four-dimensions, three-dimensions, and two-dimensions,''
Int.\ J.\ Mod.\ Phys.\ D {\bf 5} (1996) 763
{\tt  [arXiv:hep-th/9611024]}.

\bibitem{Vasiliev:1999ba}
M.A.~Vasiliev,
``Higher spin gauge theories: Star product and AdS space,"
in: {\it The many faces of the superworld}, Yuri Golfand Memorial Volume
eds Y.~Golfand and M.A.~Shifman, World Scientific (1999)
533
{\tt [arXiv:hep-th/9910096]}.

\bibitem{Gaberdiel:2012uj} 
M.R.~Gaberdiel and R.~Gopakumar,
``Minimal model holography,''
J.\ Phys.\ A {\bf 46} (2013) 214002
{\tt [arXiv:1207.6697 [hep-th]]}.

\bibitem{Gaberdiel:2010pz}
M.R.~Gaberdiel and R.~Gopakumar,
 ``An AdS$_3$ dual for minimal model CFTs,''
Phys.\ Rev.\ D {\bf 83} (2011) 066007
 {\tt [arXiv:1011.2986 [hep-th]]}.

\bibitem{Sundborg:2000wp}
B.~Sundborg,
``Stringy gravity, interacting tensionless strings and massless higher spins,"
Nucl.\ Phys.\ Proc.\ Suppl.\  {\bf 102} (2001) 113
{\tt [arXiv:hep-th/0103247]}.

\bibitem{Witten}
E.~Witten, talk at the John Schwarz 60-th birthday symposium (Nov. 2001), \newline
{\tt http://theory.caltech.edu/jhs60/witten/1.html}.

\bibitem{Mikhailov:2002bp}
A.~Mikhailov,
``Notes on higher spin symmetries,"
{\tt arXiv:hep-th/0201019}.

\bibitem{Sagnotti:2011qp}
A.~Sagnotti,
``Notes on strings and higher spins,''
J.\ Phys.\ A {\bf 46} (2013) 214006
{\tt [arXiv:1112.4285 [hep-th]]}.

\bibitem{Chang:2012kt}
C.-M.~Chang, S.~Minwalla, T.~Sharma and X.~Yin,
``ABJ triality: from higher spin fields to strings,''
J.\ Phys.\ A {\bf 46} (2013) 214009
{\tt [arXiv:1207.4485 [hep-th]]}.

\bibitem{Gaberdiel:2013vva}
M.R.~Gaberdiel and R.~Gopakumar,
``Large $\mathcal{N}=4$ holography,''
 JHEP {\bf 1309} (2013) 036
{\tt  [arXiv:1305.4181 [hep-th]]}.

\bibitem{Gukov:2004ym}
S.~Gukov, E.~Martinec, G.W.~Moore and A.~Strominger,
``The search for a holographic dual to AdS$_3 \times {\rm S}^3 \times {\rm S}^3 \times {\rm S}^1$,''
Adv.\ Theor.\ Math.\ Phys.\  {\bf 9} (2005) 435
{\tt [arXiv:hep-th/0403090]}.

\bibitem{Tong:2014yna}
D.~Tong,
``The holographic Dual of AdS$_3 \times {\rm S}^3 \times {\rm S}^3 \times {\rm S}^1$
{\tt arXiv:1402.5135 [hep-th]}.

\bibitem{OhlssonSax:2011ms}
O.~Ohlsson Sax and B.~Stefanski, Jr.,
``Integrability, spin-chains and the AdS3/CFT2 correspondence,''
JHEP {\bf 1108} (2011) 029
{\tt [arXiv:1106.2558 [hep-th]]}.


\bibitem{Borsato:2012ss}
R.~Borsato, O.~Ohlsson Sax and A.~Sfondrini,
``All-loop Bethe ansatz equations for AdS3/CFT2,''
JHEP {\bf 1304} (2013) 116
{\tt [arXiv:1212.0505 [hep-th]]}.


\bibitem{Candu:2013fta}
C.~Candu and C.~Vollenweider,
``On the coset duals of extended higher spin theories,''
{\tt arXiv:1312.5240 [hep-th]}.

\bibitem{Creutzig:2013tja}
T.~Creutzig, Y.~Hikida and P.~B.~Ronne,
``Extended higher spin holography and Grassmannian models,''
JHEP {\bf 1311} (2013) 038
{\tt [arXiv:1306.0466 [hep-th]]}.

\bibitem{Gaberdiel:2014yla}
M.R.~Gaberdiel and C.~Peng,
``The symmetry of large N=4 holography,''
{\tt arXiv:1403.2396 [hep-th]}.

\bibitem{Creutzig:2011fe}
 T.~Creutzig, Y.~Hikida and P.B.~Ronne,
``Higher spin AdS$_3$ supergravity and its dual CFT,''
JHEP {\bf 1202} (2012) 109
{\tt  [arXiv:1111.2139 [hep-th]]}.

\bibitem{Candu:2012jq}
 C.~Candu and M.R.~Gaberdiel,
``Supersymmetric holography on $AdS_3$,''
JHEP {\bf 1309} (2013) 071
{\tt [arXiv:1203.1939 [hep-th]]}.

\bibitem{triality}
M.R.~Gaberdiel and R.~Gopakumar,
``Triality in minimal model holography,''
JHEP {\bf 1207} (2012) 127
{\tt  [arXiv:1205.2472 [hep-th]]}.

\bibitem{Candu:2012tr}
C.~Candu and M.R.~Gaberdiel,
``Duality in N=2 Minimal Model Holography,''
JHEP {\bf 1302} (2013) 070
{\tt [arXiv:1207.6646 [hep-th]]}.

\bibitem{goddard}
P.~Goddard and A.~Schwimmer,
``Factoring out free fermions and superconformal algebras,''
Phys.\ Lett.\ B {\bf 214} (1988) 209.

\bibitem{Fradkin:1992km}
E.S.~Fradkin and V.Y.~Linetsky,
``Classification of superconformal and quasi-superconformal algebras in two-dimensions,''
Phys.\ Lett.\ B {\bf 291} (1992) 71.

\bibitem{henneaux}
M.~Henneaux, L.~Maoz and A.~Schwimmer,
``Asymptotic dynamics and asymptotic symmetries of three-dimensional extended AdS supergravity,''
Annals Phys.\  {\bf 282} (2000) 31
{\tt  [arXiv:hep-th/9910013]}.

\bibitem{jeugt}
J. van der Jeugt, ``Irreducible representations of the exceptional Lie superalgebras $D(2,1;\alpha)$,''
J.\ Math.\ Phys.\ {\bf 26} (1985) 913.

\bibitem{evenhol}
C.~Candu, M.R.~Gaberdiel, M.~Kelm and C.~Vollenweider,
``Even spin minimal model holography,''
JHEP {\bf 1301} (2013) 185
{\tt  [arXiv:1211.3113 [hep-th]]}.

\bibitem{Candu:2013uya}
  C.~Candu and C.~Vollenweider,
``The $\mathcal{N} =$ 1 algebra $\mathcal{W}_\infty[\mu]$ and its truncations,''
JHEP {\bf 1311} (2013) 032
{\tt  [arXiv:1305.0013 [hep-th]]}.

\bibitem{Beccaria:2013wqa}
M.~Beccaria, C.~Candu, M.R.~Gaberdiel and M.~Groher,
``$\mathcal{N}=1$ extension of minimal model holography,''
{\tt  arXiv:1305.1048 [hep-th]}.

\bibitem{wrep}
  P.~Bouwknegt and K.~Schoutens,
``$\mathcal{W}$ symmetry in conformal field theory,''
Phys.\ Rept.\  {\bf 223} (1993) 183
{\tt  [arXiv:hep-th/9210010]}.

\bibitem{Blumenhagen:1990jv}
R.~Blumenhagen, M.~Flohr, A.~Kliem, W.~Nahm, A.~Recknagel and R.~Varnhagen,
``$\mathcal{W}$ algebras with two and three generators,''
Nucl.\ Phys.\ B {\bf 361} (1991) 255.

\bibitem{Thielemans:1994er}
K.~Thielemans,
``An Algorithmic approach to operator product expansions, $\mathcal{W}$ algebras and $\mathcal{W}$ strings,''
{\tt  arXiv:hep-th/9506159}.

\bibitem{Thielemans:1991uw}
K.~Thielemans,
``A Mathematica package for computing operator product expansions,''
Int.\ J.\ Mod.\ Phys.\ C {\bf 2} (1991) 787.

\bibitem{Ahn:2013oya}
C.~Ahn,
``Higher spin currents in Wolf space. Part I,''
JHEP {\bf 1403} (2014) 091
{\tt [arXiv:1311.6205 [hep-th]]}.

\bibitem{King:1971rs}
R.C.~King,
``Modification rules and products of irreducible representations of the unitary, orthogonal, and symplectic groups,''
J.\ Math.\ Phys.\  {\bf 12} (1971) 1588.

\bibitem{Sevrin:1988ew}
  A.~Sevrin, W.~Troost and A.~Van Proeyen,
``Superconformal algebras in two-dimensions with N=4,''
Phys.\ Lett.\ B {\bf 208} (1988) 447.

\bibitem{Schoutens:1988ig}
K.~Schoutens,
``O(n) extended superconformal field theory in superspace,''
Nucl.\ Phys.\ B {\bf 295} (1988) 634.

\bibitem{Spindel:1988sr}
P.~Spindel, A.~Sevrin, W.~Troost and A.~Van Proeyen,
``Extended supersymmetric sigma models on group manifolds. 1. The complex structures,''
Nucl.\ Phys.\ B {\bf 308} (1988)  662.

\bibitem{VanProeyen:1989me}
A.~Van Proeyen,
``Realizations of N=4 superconformal algebras on Wolf spaces,''
Class.\ Quant.\ Grav.\  {\bf 6} (1989) 1501.

\bibitem{Sevrin:1989ce}
A.~Sevrin and G.~Theodoridis,
``N=4 superconformal coset theories,''
Nucl.\ Phys.\ B {\bf 332} (1990) 380.

\bibitem{Ivanov:1992rt}
E.~Ivanov, S.~Krivonos, and V.~Leviant,
``N=3 and N=4 superconformal WZNW   sigma models in superspace. 2: The N=4 case,"  
Int.\ J.\ Mod.\ Phys.\  A {\bf  7} (1992) 287.

\bibitem{Ivanov:1989qs}
E.~Ivanov, S.~Krivonos, and V.~Leviant, 
``N=3 and N=4 superconformal WZNW   sigma models in superspace. 1. General formalism and N=3 case,"  
Int.\ J.\ Mod.\ Phys.\ A {\bf 6} (1991) 2147.

\bibitem{Nagi:2004wb}
J.~Nagi, 
``On extensions of superconformal algebras,"  
J.\ Math.\ Phys.\  {\bf 46} (2005) 042308 
{\tt  [arXiv:hep-th/0412061]}.




	
\end{thebibliography}
\end{document}